\documentclass{article}
\usepackage{graphicx} 
\usepackage{subfigure}
\usepackage{bm}
\usepackage{amsmath}
\usepackage{amssymb}
\usepackage{xcolor}
\usepackage{amsthm} 
\usepackage{mathtools}
\usepackage{authblk}
\usepackage{hyperref}
\usepackage{natbib}
\usepackage{orcidlink}

\title{{Neural operator accelerated atomistic to continuum concurrent multiscale simulations of viscoelasticity}\footnote{Notice: This manuscript has been coauthored by UT-Battelle, LLC, under Contract No. DE-AC0500OR22725 with the U.S. Department of Energy. The United States Government retains and the publisher, by accepting the article for publication, acknowledges that the United States Government retains a non-exclusive, paid-up, irrevocable, world-wide license to publish or reproduce the published form of this manuscript, or allow others to do so, for the United States Government purposes. The Department of Energy will provide public access to these results of federally sponsored research in accordance with the DOE Public Access Plan (\href{http://energy.gov/downloads/doe-public-access-plan}{http://energy.gov/downloads/doe-public-access-plan}).}}
\author[1]{Tanvir Sohail\orcidlink{0000-0001-7567-6417}}
\author[2]{Burigede Liu\orcidlink{0000-0002-6518-3368}}
\author[1*]{Swarnava Ghosh\orcidlink{0000-0003-3800-5264}}
\affil[1]{National Center for Computational Sciences, Oak Ridge National Laboratory, TN 37830}
\affil[2]{Department of Engineering, University of Cambridge, Cambridge, CB2 1PZ, UK}
\affil[*]{Email: ghoshs@ornl.gov}
\begin{document}

\maketitle
\begin{abstract}
We present a neural-operator-accelerated concurrent multiscale framework that couples atomistic simulations with continuum finite-element analysis for history-dependent materials, thereby making atomistic-continuum multiscale simulations of viscoelastic materials tractable. The approach replaces direct molecular dynamics (MD) evaluation of the constitutive response with a Recurrent Neural Operator (RNO) surrogate trained on atomistic simulations. The surrogate learns the strain-history-to-stress operator from molecular dynamics simulations and provides a discretization-independent approximation of the atomistic constitutive mapping, enabling efficient evaluation of stresses and latent internal variables at each quadrature point. The framework is implemented within an explicit finite-element solver, where the constitutive update reduces to inexpensive operator evaluations rather than repeated MD solves. Memory effects are represented through learned internal states, and transfer learning across temperature enables the surrogate to capture thermally dependent viscoelastic behavior. The method is assessed using polyurea through cyclic loading, Taylor impact, and plate impact simulations and compared with an experimentally calibrated viscoelastic polyurea model and a Johnson–Cook model. The neural-operator surrogate reproduces correct viscoelastic response while enabling atomistically informed dynamic simulations at scales that are not tractable with direct MD–FEM coupling.

\end{abstract}
\section{Introduction}



The macroscopic response of materials arises from the interaction of multiple mechanisms that play out over a hierarchy of length and time scales \cite{phillips2001crystals}. Over the course of the last few decades, multiscale modeling frameworks have been developed to address the challenges of modeling complex material behavior with high predictive accuracy \cite{van2020roadmap}. Multiscale modeling is commonly classified into hierarchical and concurrent approaches or a combination of both \cite{elmasry2023comparative}. In a hierarchical multiscale method, the different scales are weakly coupled, and information is passed in one direction from one scale to another, typically from the fine-scale to the coarse-scale. For example, fine-scale models such as atomistic simulations or mesoscale calculations are used to obtain effective material properties and constitutive relations, such as elastic moduli, yield criteria, transport coefficients, and internal length scales. These quantities are subsequently supplied as inputs to a coarse-scale continuum model, which is solved independently, without real-time feedback from the fine scale. Owing to this sequential coupling, hierarchical approaches are computationally efficient; however, they can have less predictive accuracy for materials exhibiting strong cross-scale interactions, evolving microstructures, or localization phenomena that depend sensitively on the macroscopic loading path \cite{fish2010multiscale}.

In contrast, concurrent multiscale methods couple multiple scales within a single, unified simulation, enabling explicit resolution of nonlocal effects, defect–microstructure interactions, and complex phenomena such as microstructural evolution and rate-dependent phenomena. However, this increased fidelity comes at the cost of substantially higher computational cost and algorithmic complexity \cite{fish2010multiscale}.

Over the past few decades, several methods enabling concurrent multiscale modeling have been proposed and developed. Concurrent multiscale methods that couple atomistic and continuum descriptions include the quasicontinuum (QC) method
\cite{tadmor1996quasicontinuum,shenoy1999adaptive,miller2002quasicontinuum},
the bridging scale
\cite{wagner2003coupling}
and bridging domain methods
\cite{xiao2004bridging},
and the concurrent atomistic--continuum (CAC) formulation
\cite{xiong2011coarse,xu2015quasistatic}.
These methods simultaneously resolve atomistic and continuum length scales. In the quasicontinuum approach, full atomistic resolution is retained in regions of large deformation gradients, such as near defects and crack tips, while smoothly deformed regions are treated using a continuum approximation obtained through interpolation of atomic positions. In the bridging scale method, the displacement field is decomposed into coarse-scale continuum and fine-scale atomistic components, enabling concurrent coupling through scale separation operators while retaining full atomistic resolution in selected regions. The bridging domain formalism introduces a geometric overlap region in which atomistic and continuum descriptions coexist, with kinematic compatibility and equilibrium enforced through constraint-based coupling or energy interpolation within the overlap. In the concurrent atomistic–continuum method, the atomistic and continuum subdomains coexist within a single physical domain, and are coupled by enforcing consistency through common balance laws, without needing overlap regions, projection operators, or reduced atomistic representations. These methods assume the presence of an underlying lattice and are therefore typically applied to crystalline materials; as a result, their applicability to study disordered materials such as polymers, rate-dependent phenomena, materials with memory, and viscoelastic behavior remains limited.


{The FE$^2$ method \cite{feyel2000fe2,feyel2003multilevel} is a concurrent multiscale approach in which both scales are treated as continua and coupled via computational homogenization, with the macroscopic response obtained from microscale boundary value problems posed on representative volume elements at each integration point.}
and consistency of energy between the two scales is ensured by the Hill-Mandel condition \cite{hill1963elastic,mandel1964contribution}. This method enables the modeling of complex nonlinear material responses, including damage and plasticity, in heterogeneous media. An essential constituent of the FE$^2$ approach stems from the theory of homogenization \cite{tartar2009general}, where balance laws at the macro or coarse scale are solved using constitutive relations determined by solving several unit cell problems defined at the micro or fine scale. In practice, implementing this is prohibitively computationally expensive because of the repeated solution of unit cell problems.

Several important distinctions exist between the quasicontinuum, bridging scale, bridging domain, and concurrent atomistic–continuum (CAC) methods, and the FE$^2$ approach. First, in FE$^2$, the governing physics at both the fine and coarse scales are formulated entirely within a continuum framework. Second, FE$^2$ assumes a clear separation of scales, whereas the other methods concurrently resolve atomistic and continuum length scales within a single simulation. Finally, while the quasicontinuum, bridging scale, bridging domain, and CAC methods are primarily suited to crystalline materials, the FE$^2$ approach has been successfully applied to non-crystalline materials \cite{tikarrouchine2018three,tikarrouchine2021non,oukfif2024multi}. 

{The heterogeneous multiscale method (HMM)~\cite{e2003hmm,vassaux2019hmm} is an atomistic–continuum concurrent multiscale approach in which a fine-scale model is evaluated at each macroscopic quadrature point to provide the constitutive information required at the coarse scale under an assumed scale separation. This necessitates on-the-fly solution of the fine-scale problem during the macroscopic simulation, leading to a substantial computational burden that, for many problems, renders the approach impractical.} {In addition, multiscale frameworks specifically designed for polymers, such as the Capriccio approach \cite{pfallerarlequin,zhouIJNME}, introduce an atomistic-to-continuum concurrent coupling tailored to capture polymer chain dynamics within an overlap region, while subsequent developments \cite{adityaMSMS,aditya2024asme,norouziJCTC} extend the framework to polymer networks, interphases, and dynamic loading. These methods are particularly well suited for problems in which localized atomistic features, such as crack tips, interphases, or free surfaces, require explicit resolution within an otherwise continuum domain.} 

In this work, we focus on viscoelastic materials, with particular emphasis on polymeric systems. In these materials, the mechanical response depends not only on the current state of deformation but also on its history and the rate at which deformation is applied (\cite{christensen2013theory}). These materials also exhibit several interesting phenomena, including creep, stress relaxation, and hysteresis \cite{anand2020continuum}. Multiscale modeling of viscoelastic materials is challenging because of complexities arising from homogenizing history-dependent materials. Specifically, heterogeneous distributions of strain and strain rate within the unit cell give rise to interactions among deformation mechanisms characterized by distinct intrinsic time scales; and as a consequence, the effective history dependence emerging at the macroscopic level can differ substantially from the local constitutive behavior \cite{liu2023learning,sanchez1978certains,francfort1986homogenization}. 

The challenges in multiscale modeling of materials have motivated research aimed at incorporating machine learning to accelerate multiscale simulations \cite{anand2023exploiting}. Lefik and co-workers \cite{lefik2009artificial} used artificial neural networks for multiscale simulations of composite materials, demonstrating that machine learning can be used to recover stress-strain relations, estimate the state of yielding, and define homogenized properties of unit cells with improved computational efficiency. Ghavamian and co-workers \cite{ghavamian2019accelerating} used a recurrent neural network (RNN) with long short-term memory to describe the effective plastic behavior of representative volume elements, which can be used for multiscale simulations and demonstrated their framework on academic examples. Deep learning methods can also accurately and efficiently predict material plasticity \cite{mozaffar2019deep}. Specifically, a recurrent neural network with gated recurrent units can approximate the homogenized response of composites with elastic inclusions in a rate-independent plastic matrix. Recurrent neural networks can also approximate the response of the unit cell of a composite of circular elastic particles in an isotropic matrix where the matrix is rate-independent with J2 plasticity and isotropic hardening \cite{wu2020recurrent,wu2022recurrent}. Furthermore, the trained network is used as a surrogate in macroscopic simulations. Self-consistent recurrent neural networks \cite{bonatti2022importance} and long short-term memory neural networks with proper orthogonal decomposition \cite{im2021surrogate} were used as a surrogate model of elasto-plastic solids in finite element simulations. Logarzo and co-workers \cite{logarzo2021smart} developed a generic, microstructure-independent framework for computational homogenization using machine learning for concurrent multiscale models that can handle non-linearities and path dependencies. In linear elastic fiber-reinforced composites, a deep convolutional neural network is used to map the two-dimensional stress fields to the fiber orientation \cite{gupta2023accelerated}. Machine learning of electronic fields \cite{teh2021machine} can be used to accelerate large scale first principles calculations of materials \cite{ghosh2017sparc1,ghosh2017sparc2,xu2021sparc} and provide insight into interesting phenomena such as strain engineering \cite{ghosh2019electronic,ghosh2025strain}, defect interactions \cite{ghosh2022spectral}, microstructure evolution \cite{ghosh2020influence,ghosh2021precipitation}. We refer the reader to \cite{mirkhalaf2024micromechanics} for a brief review of the application of machine learning for multiscale modeling of composites. 

Atomistic simulations have also been used to develop machine-learned surrogates for FEM simulations. Hasan and co-workers \cite{hasan2025machine} a combined classification-regression neural network surrogate model bridging atomistic insights with continuum scale finite element simulations for Al-SiC nanocomposites. Machine learning regression and classification models for Ti-TiB$_2$ metal ceramic composite were also developed in \cite{xiao2021machine}. We note that these composites are linear elastic, and these machine learning methods employed in these simulations (i.e., in Refs. \cite{hasan2025machine,xiao2021machine}) cannot be applied to capturing the rate-dependent response of materials.

The Recurrent Neural Operator (RNO) \cite{liu2023learning,bhattacharya2023learning} addresses the challenge of developing surrogates for viscoelastic materials by learning the mapping from strain, strain rate, and stress. Furthermore, this operator is resolution-independent and therefore provides a good approximation for discretizations beyond those used for training. In Ref. \cite{liu2023learning}, solutions of PDEs describing the history-dependent response of mesoscale microstructures are used to train the RNO and demonstrate the accuracy and efficiency of the surrogate using FE$^2$ simulations of viscoelastic composites and elasto-viscoplastic polycrystals. We also refer the reader to Ref \cite{bhattacharya2023learning} for a detailed theoretical exposition of the recurrent neural operator. The recurrent neural operator has also been extended to develop surrogates for materials exhibiting temperature memory \cite{hollenweger2026temperature}. It has also been applied for modeling viscoelastic metamaterials \cite{zhang2024iterated} and reactive flow through porous microstructures \cite{karimi2024learning}. 

Internal variable theories in continuum mechanics \cite{rice1971inelastic} postulate that the entire prior deformation history of a body can be represented by a finite set of internal variables that evolve with the deformation. Consequently, the local stress at any given time is determined solely by the current deformation gradient and the values of the internal variables at that material point. The evolution of the internal variables is governed by constitutive evolution laws that depend locally on the deformation gradient and the internal variables themselves.

Atomistic simulations play a critical role in elucidating the constitutive response of viscoelastic materials such as polymers because they directly capture the microscopic chain dynamics that govern time- and rate-dependent behavior without empirical assumptions \cite{likhtman2007linear}.

The central contribution of this work is the development of a concurrent atomistic–continuum multiscale framework for history-dependent materials, wherein fine-scale atomistic information is systematically upscaled via a learned surrogate model. Atomistic simulations of polyurea-based polymer composites are used to construct training data, from which a recurrent neural operator is trained to represent the underlying viscoelastic constitutive response. The learned operator is then embedded within the multiscale formulation, enabling predictive simulations of polyurea composites under static and dynamic loading. This approach eliminates the need for repeated atomistic solves at quadrature points and renders atomistic-to-continuum simulations of viscoelasticity computationally tractable. {In the multiscale simulations, the influence of slow processes are encoded implicitly through latent internal variables learned from short-time molecular dynamics, which provide a reduced-order representation of unresolved relaxation mechanisms governing the macroscopic history-dependent response.}

{We also clarify the novelty of the present work relative to recent studies that employed Recurrent Neural Operators (RNOs) as continuum surrogates for viscoelastic microstructures \cite{liu2023learning,bhattacharya2023learning,shenCS}. The current contribution differs in three principal respects. First, the RNO is trained directly on molecular-dynamics trajectories rather than on solutions of continuum unit-cell partial differential equations. Consequently, the learned surrogate inherits chain-segment dynamics, hydrogen-bond relaxation, and hard/soft microphase interactions directly from interatomic forces, without introducing an empirical microscale constitutive model. Second, the trained operator is embedded within a production-scale explicit finite-element solver, demonstrated and validated on three representative large-deformation dynamic problems involving contact, stress-wave propagation, and thermomechanical coupling, at scales that are computationally prohibitive for direct concurrent MD–FEM simulations. Third, we employ transfer learning across temperature, whereby an operator trained at 300K is fine-tuned for 400K and 500K using substantially less additional data, thereby enabling a temperature-dependent constitutive surrogate at a fraction of the molecular-dynamics cost required for independent retraining.}

We organize the remainder of this paper as follows. In Section \ref{sec:Theory}, we first give an overview of the recurrent neural operator for mapping stress to deformation gradient and rate of change of deformation gradient, followed by a discussion on sampling of the data and training. Next, we provide an overview of the accelerated multiscale formulation. In Section \ref{Sec:Learning}, we first discuss the details of three-dimensional atomistic simulations of polymers, followed by a discussion on the atomistic surrogate. In Section \ref{sec:multiscale}, we present the results of multiscale simulations of polyurea composites using the atomistic surrogate in finite element simulations. In Section \ref{sec:efficiency}, we present timings and discuss the computational efficiency. Finally, in Section \ref{sec:conc} we present concluding remarks and future directions of research.

\section{Theory and methods}\label{sec:Theory}
\subsection{Background of Recurrent Neural Operator}\label{sec:background}
In this section, we provide an overview of the Recurrent Neural Operator (RNO) \cite{liu2023learning,bhattacharya2023learning}. Viscoelastic materials exhibit history dependence where the stress at a given time does not depend solely on the instantaneous deformation but on the entire prior loading path. Additionally, stress can also depend on temperature. In view of this, the effective constitutive relation is the map
\begin{equation}
\varPhi:\ \{\bm{F}(\tau),\bm{\dot{F}}(\tau),\theta(\tau):\tau\in(0,t)\}\ \mapsto\ \bm{\sigma}(t),
\label{eq:psi_operator}
\end{equation}
where $\tau$ is time, $\bm{F}(\tau)$ is the macroscopic deformation gradient, $\bm{\dot{{F}}}(\tau)$ is it's temporal derivative, $\theta(\tau)$ is temperature, and $\bm{\sigma}(\tau)$ is the Cauchy stress tensor.

In a multiscale simulation, the mapping in Equation \eqref{eq:psi_operator} is obtained from the solution of fine-scale unit cell problems, where the deformation gradient, its time derivative, and the temperature enter as boundary conditions. {Because this microscale problem must be solved at every quadrature point of the coarse-scale model and at every time step \cite{e2003hmm,vassaux2019hmm}}, the resulting computational cost becomes prohibitively large, and in many cases renders the direct approach intractable. The recurrent neural operator approximates this map by learning from the solutions of a set of unit cell problems. 

RNO approximates this map using the form
\begin{subequations}
\label{eq:rno_continuous}
\begin{align}
\bm{\sigma}(t) &= \bm{\varphi}\!\left(\bm{F}(t),\bm{\dot{F}}(t),\theta(t),\bm{\xi}(t)\right),\\
\dot{\bm{\xi}}(t) &= \bm{\beta}\!\left(\bm{F}(t),\bm{\dot{F}}(t),\theta(t),\bm{\xi}(t)\right),
\end{align}
\end{subequations}
where $\bm{\xi}(t)=\{\xi_1(t),\xi_2(t),\ldots,\xi_{k}(t)\}$ is a vector of latent internal variables and $\bm{\varphi}$ and {$\bm{\beta}$} are neural networks. The internal variables are used to describe inelastic behavior \cite{rice1971inelastic} and not prescribed \emph{a priori}, but are discovered through training so that the pair $(\bm{\varphi},\bm{\beta})$ reproduces the fine-scale constitutive response. Equation \eqref{eq:rno_continuous} yields a time-continuous system whose solution defines a mapping from a deformation history and temperature to a stress history, thereby approximating $\varPhi$ (Eq. \eqref{eq:psi_operator}). As the functions $\bm{\varphi}$ and $\bm{\beta}$ are applied repeatedly in time, they are hence recurrent. {Upon discretizing in time with time step $(\Delta t)_n = t_n-t_{n-1}$, we employ an explicit forward Euler update for the evolution of the internal variables. The kinematic inputs $(\bm{F})_{n},(\bm{\dot{F}})_{n}$ are evaluated at the current time $t_n$, as these quantities are provided directly by the explicit finite-element solver at each increment. Accordingly, the discrete form of Equation \ref{eq:rno_continuous} at time $t_n$ is written as follows.}

\begin{subequations}
\label{eq:rno_discrete}
\begin{align}
(\bm{\sigma})_{n} &= \bm{\varphi}\!\left((\bm{F})_{n},(\bm{\dot{F}})_{n},\theta,(\bm{\xi})_{n}\right)
\label{eq:rno_discrete_a} \\
(\bm{\xi})_{n} &= (\bm{\xi})_{n-1} + (\Delta t)_{n}\,\bm{\beta}\!\left((\bm{F})_{n},(\bm{\dot{F}})_{n},\theta,(\bm{\xi})_{n-1}\right)
\label{eq:rno_discrete_b}
\end{align}
\end{subequations}
In our work, we use transfer learning \cite{goodfellow2016deep} to approximate the stress at multiple temperatures. RNO has also been extended to materials showing temperature memory \cite{hollenweger2026temperature}. Additionally, the discretized RNO (Equation \ref{eq:rno_discrete}) resembles a recurrent neural network, but it differs from gated architectures (e.g., LSTM/GRU) in that memory is specified explicitly by latent internal variables governed by an evolution law. Another advantage of RNO is its explicit continuous-time structure, which enables discretization independence. In practice, both training and application involve time-discretized data, potentially at different time resolutions. In our work, we choose time steps that are independent of $n$ (i.e., $(\Delta t)_{n}=\Delta t$). {The model is trained end-to-end using backpropagation through time to minimize the discrepancy between predicted and atomistically computed stresses over the training trajectories.}

In the RNO formulation, the internal variables $\bm{\xi}(t) = \{\xi_1(t), \xi_2(t), \ldots, \xi_k(t)\}$ are calculated during the time-stepping process. However, the number of internal variables $k$ must be specified a priori. In practice, this is determined by training separate networks for different values of $k$, and then examining the resulting error as $k$ increases to assess convergence.

\begin{figure}[h!]
\centering
{\includegraphics[keepaspectratio=true,width=0.75\textwidth]{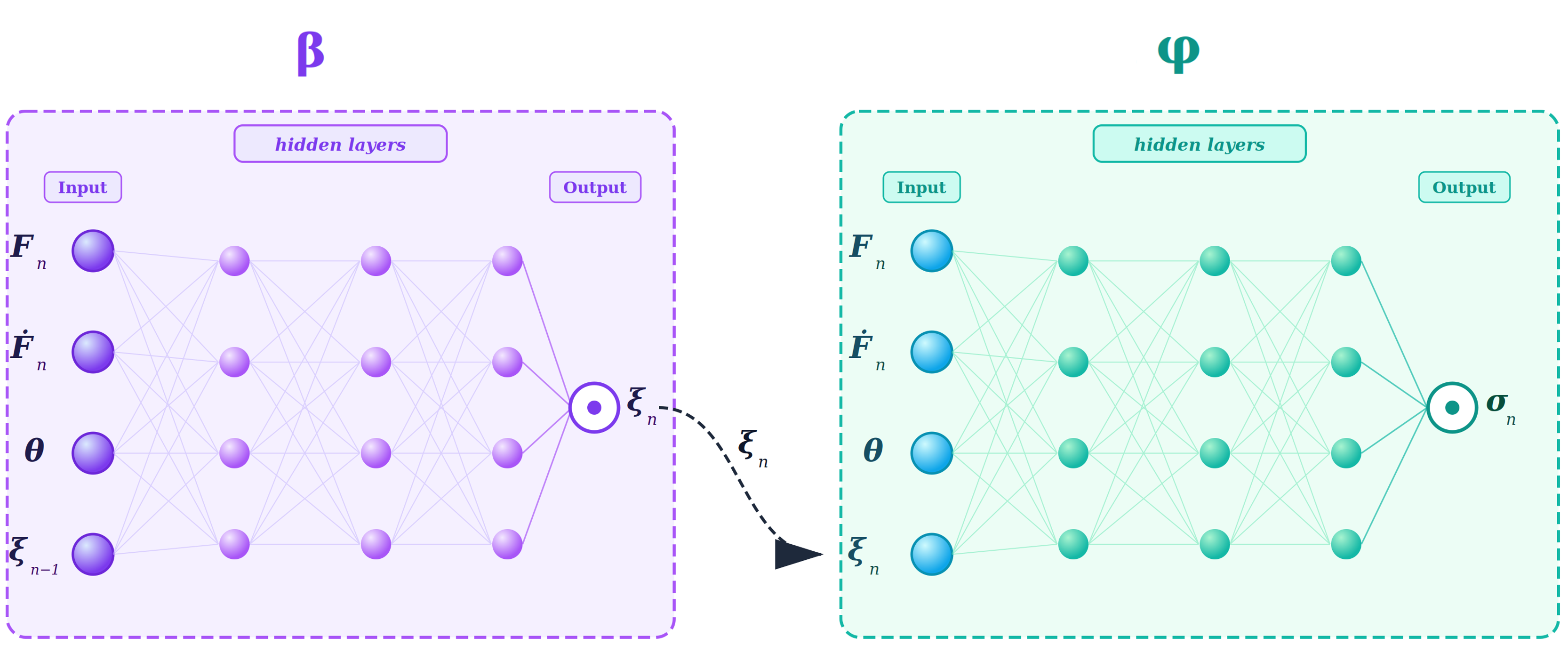}}
{\caption{Neural network diagram of discretized RNO at the $n^{th}$ time step.}\label{Fig:NN}}
\end{figure}

{We use the Adam algorithm \cite{adam2014method} for optimizing the model parameters, and the scaled exponential linear unit (SELU) \cite{SELU} as the non-linear activation function, learning rate of $10^{-5}$, batch size of $32$, and $2$ hidden layers with $100$ neurons in each layer. The loss function is defined as the relative $L^2$ error between the predicted and true stress components over the entire trajectories.} Figure \ref{Fig:NN} shows a schematic of the discrete version of the recurrent neural operator using a network diagram.

To sample the data, the deformation gradient is varied smoothly in time but changes direction arbitrarily. This is given by
\begin{equation}\label{Eq:deformationpath}
(\bm{{F}})_m = (\bm{{F}})_{m-1} + \bm{\nu}_m \, {F}_{\max} \sqrt{(\Delta t)_m},
\end{equation}
where $\bm{\nu}_m$ is a tensor whose independent components are sampled
from a Rademacher distribution with values $+1$ or $-1$, ${F}_{\max}$ is the maximum magnitude of deformation gradient. The total time interval $[0,T]$ is partitioned into $M$ subintervals $0=t_0<t_1<\dots<t_M=T$, and $\bm{F}(t)$ is obtained from $(t_m,(\bm{F})_m)$ by cubic Hermite interpolation.

\subsection{Neural operator accelerated multiscale formulation}\label{sec:AtCont}
We motivate our formalism in the spirit of the FE$^2$ method \cite{feyel2003multilevel}, where the constitutive law is derived from microscale finite element simulations, followed by homogenization. In our work, the constitutive behavior of the material is defined in the microscopic atomistic scale and is determined using molecular dynamics simulations. To this end, the formulation can be separated into two distinct parts: the first being the development of the neural operator atomistic surrogate, and the second being the use of this surrogate for macroscopic simulations.

In the first part, an atomistic representative volume element (RVE) is chosen such that it correctly captures bulk properties of the material. Next, the deformation gradients are varied temporally using Equation \ref{Eq:deformationpath} for multiple strain paths, and these deformation gradients are used to apply strain on the atomistic RVEs. Next, for each of these atomistic RVEs, molecular dynamic simulations using the NVT ensemble are performed, and the Cauchy stress of the atomistic RVE is calculated using the virial formula\cite{tadmor2011modeling}
\begin{equation}
\boldsymbol{\sigma}_{at} = \frac{1}{|\Omega|} \left\langle \sum_{i} m_i \, \mathbf{v}_i \otimes \mathbf{v}_i + \sum_{i<j} \mathbf{r}_{ij} \otimes \mathbf{f}_{ij} \right\rangle,
\end{equation}
where $|\Omega|$ is the volume of the deformed RVE, $m_i$ is the mass and $\mathbf{v}_i$ is the velocity of an atom with index $i$, $\mathbf{r}_{ij}$ is the pairwise distance, and $\mathbf{f}_{ij}$ is the pair-wise force. 

Once all the atomistic simulations are performed, we use the values of the stresses and Equation \ref{eq:rno_discrete} to calculate the operators $\bm{\varphi}$, $\bm{\beta}$, and the internal variables $\bm{\xi}$, which are used to define the neural operator surrogate. We store the weights of these networks for use in macroscale simulations. {We remark that the internal variables do not correspond to prescribed physical quantities, rather, they are learned during training so that the constitutive mapping $(\bm{\varphi}, \bm{\beta})$ reproduces the fine-scale response. In fact, any reparameterization of the internal variables produces an equivalent RNO, implying that the same data can produce different but equivalent choices of the internal variables depending on the initialization \cite{liu2023learning}. At each quadrature point of the macroscopic simulation and at the start of every MD training trajectory, we set
$\xi_i(0) = 0$ for $i = 1, \ldots, k$, corresponding to the undeformed, stress-free reference state; subsequent values are computed via Equation \eqref{eq:rno_discrete_b}. The number of internal variables is chosen by analyzing the error with the increase in the number of internal variables, and seeing where it converges. This has been illustrated in Section \ref{subsec:AtomisticSurrogate}.}


In our macroscale simulations, the deformation gradient, its rate of change, temperature, and internal variables are determined locally at each quadrature point through the constitutive update performed at each time increment. For each quadrature point, the solver provides the current deformation measures along with the previously stored internal variables to the material model. The internal variables are then updated through an incremental evolution law and stored for use in the next step.

In our approach, this evolution law is defined by the neural operator $\bm{\beta}$. Specifically, the internal variables are represented by a set of latent variables that encode the material history. At each increment, these variables are updated at every quadrature point as a function of the current deformation state and their previous values through the learned operator. This update is performed independently at each quadrature point, ensuring a fully local, history-dependent constitutive response. Finally, the Cauchy stress at a quadrature point and time step is determined from the internal variables, the deformation gradient, its rate of change, and temperature at that quadrature point. Note that the stress update at each quadrature point is performed locally and can therefore be efficiently parallelized on high-performance computing hardware.

\section{Atomistic simulations and surrogate}\label{Sec:Learning}

\subsection{Three-dimensional atomistic simulations of polymeric materials}\label{subsec:AtomisticSimulations}
\begin{figure}[h!]
\centering
\subfigure[chemical formula of polyurea]{\includegraphics[keepaspectratio=true,width=0.65\textwidth]{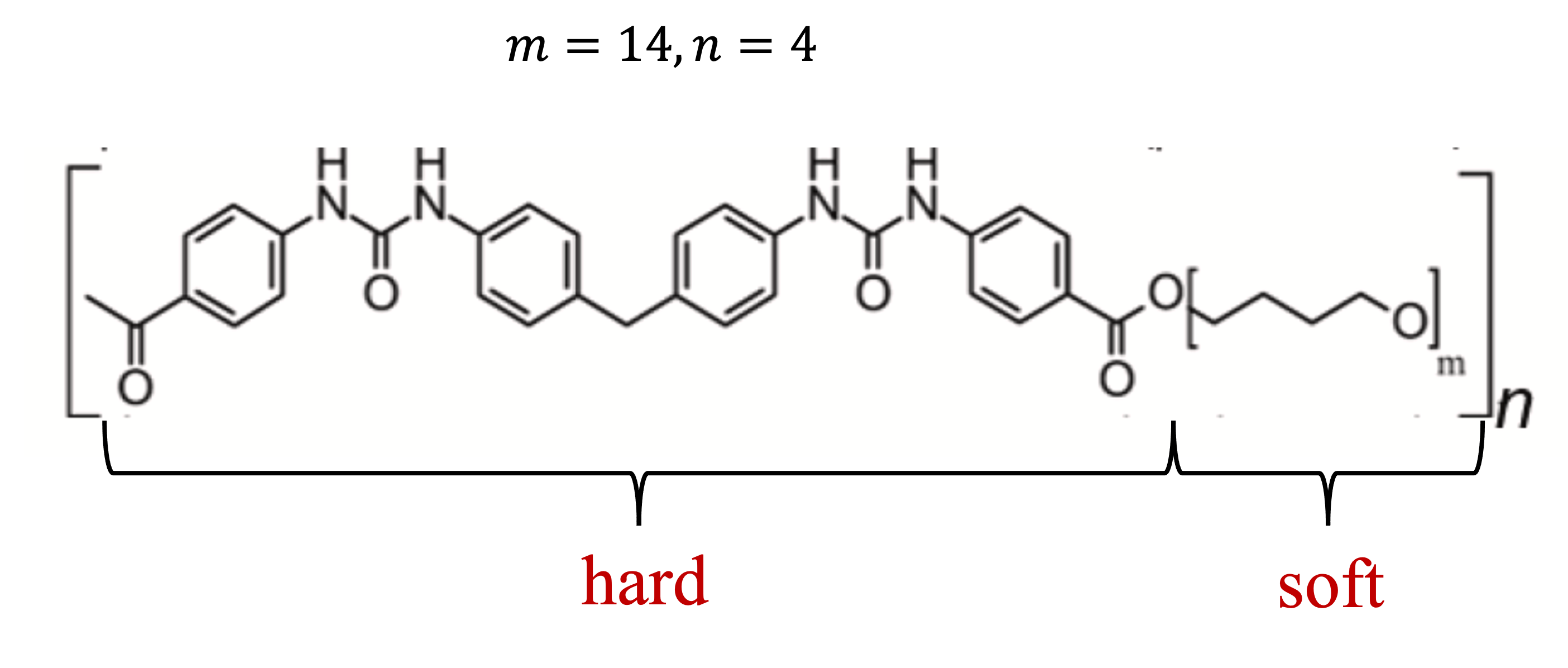}\label{Fig:PUChem}}
\subfigure[polyurea chain, unit cell and contnuum application]{\includegraphics[keepaspectratio=true,width=1.0\textwidth]{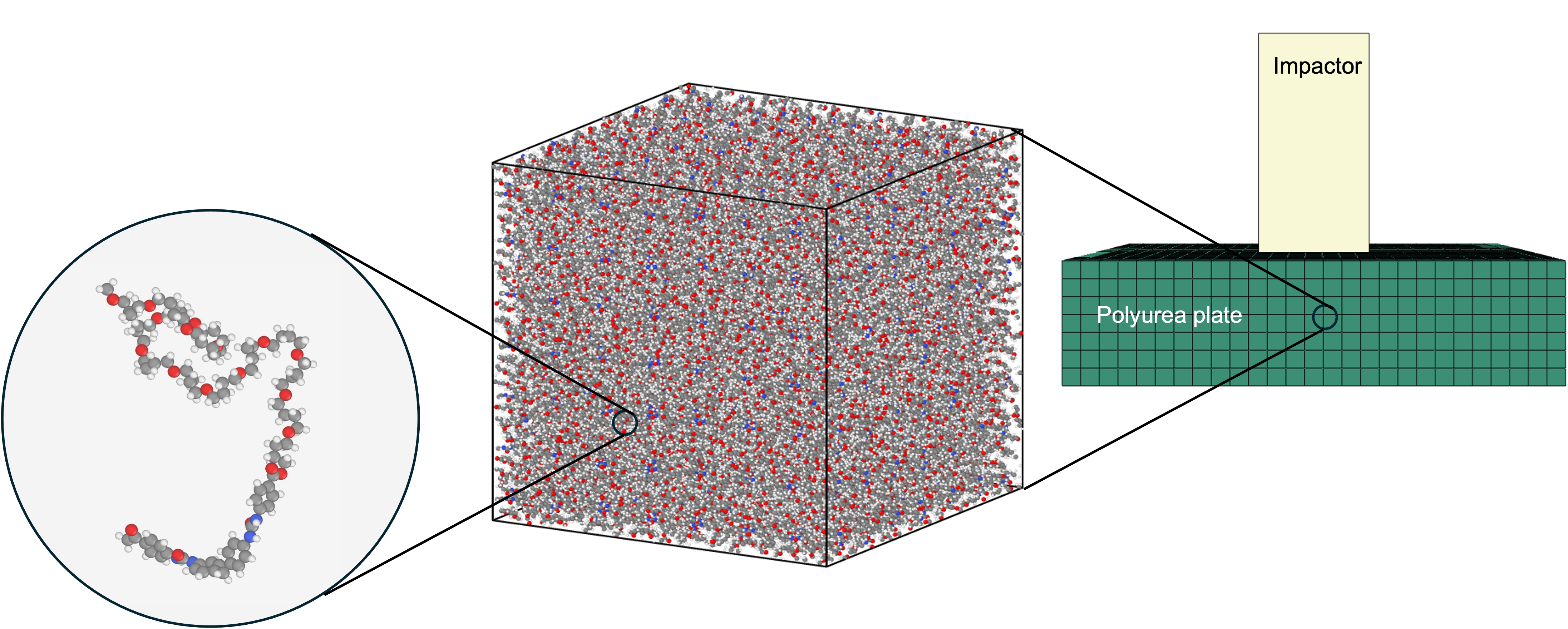}\label{Fig:PUUnitCell}}
{\caption{(a) shows the chemical formula of polyurea. Here $n$ is the number of repeating monomer units and $m$ is the number of repeating soft segments within a monomer unit. (b) shows the multiscale nature polyurea composites. The polyurea chain, atomistic representative volume element, and continuum modeling is shown.}\label{Fig:PURVE}}
\end{figure}

We used the polyurea class of polymeric materials for our simulations. The main component of polyurea is the di- or poly-isocyanate molecules containing the functional cyanate group (–NCO). This functional group exothermically reacts with amine molecules containing the functional group –NH$_2$ and produces polymers
with bonding –(NH)(CO)(NH)-, also known as urea \cite{li2018multi}. {Figure \ref{Fig:PUChem}} shows the chemical formula of a single polyurea chain with $n$ repeating monomers, where each monomer has 14 soft segments ($m = 14$) and one hard segment. In this work, we choose chains with 4 repeating monomers.

The microstructure of three-dimensional polyurea composites comprises nanoscale hydrogen-bonded hard domains dispersed within a compliant soft matrix \cite{li2018multi,zheng2022toughness}. This distinctive microphase-separated architecture gives rise to increased strain-rate sensitivity, rapid recovery under extreme loading, and high energy absorption, making polyurea particularly well suited for high strain-rate applications \cite{roland2007polyurea,clifton2016physically,jordan2022polyurea}. In Figure \ref{Fig:PUUnitCell}, we show the unit cell of a polyurea composite.

We perform molecular dynamics (MD) simulations using the Large-scale Atomic/Molecular Massively
Parallel Simulator (LAMMPS) \cite{thompson2022lammps} to characterize the mechanical behavior of amorphous polyurea across temperatures ranging from $300$ K to $500$ K. We utilize the Polymer Consistent Force Field (PCFF), a Class II polymer force field \cite{sun1994cff93,sun1998pcff}, which is well suited to capture the behavior of polymeric systems under small strain rates as well as extreme deformations \cite{dewapriya2020molecular}. In our atomistic simulations, the instantaneous atomic stress is first converted to the continuum Cauchy stress, followed by temporal averaging, which removes high-frequency vibrational fluctuations associated with atomic-scale dynamics. 

For all our MD simulations and training, we use the Oak Ridge Leadership Computing Facility Frontier supercomputer \cite{frontier2022}, and employ an ensemble simulations scheme\cite{georgiadou2024ensemble} for optimally managing the simulations. 

In our simulations, the initial configurations were amorphous and compressed to $1.12$ g/cm$^3$ at $500$ K to eliminate free volume and accelerate chain relaxation. This high-temperature treatment enhances sampling by lowering activation barriers for configurational rearrangements. The steepest descent method was used for energy minimization, followed by dynamics using the NVT ensemble for $1$ ns with the Nos\'e--Hoover thermostat, and then dynamics using the NPT ensemble for $2$ ns with a barostat relaxation time of $100$ fs. Subsequently, the system was quenched to $300$ K, $400$ K, and $500$ K using the NPT ensemble for $5$ ns with a time step of $1$ fs. This quenching enables relaxation toward thermodynamically stable morphologies at these temperatures, minimizing kinetic trapping and yielding equilibrium structures with realistic density and segmental organization.

{After the preparation and quench protocols described above, which are used solely to generate the initial equilibrated configurations (via NPT equilibration followed by NVT stabilization), each deformation trajectory is applied incrementally using $M = 1000$
strain steps per trajectory, following the incremental NVT
strain-application protocol of Heyden et al.~\cite{heyden2016all} for
all-atom molecular dynamics of polyurea. Related all-atom MD studies
of polyurea under shock loading employ analogous strain-application schemes~\cite{manav2021polyurea_shock_md,dewapriya2021quantum}. Along each deformation path, the system is evolved exclusively in the NVT ensemble.
At every strain increment, the deformation is imposed and the system
is evolved under NVT dynamics for $\tau_{\mathrm{eq}} = 1\,\mathrm{ps}$
(1000 integration steps with $\Delta t_{\mathrm{MD}} = 1\,\mathrm{fs}$)
at the target temperature using a Nos\'e--Hoover thermostat~\cite{nose1984unified}
with a damping time of $100\,\mathrm{fs}$. The Cauchy stress is then
computed from the virial stress of the supercell ~\cite{thompson2009general} at the
end of each NVT relaxation window. The choice
$\tau_{\mathrm{eq}} = 1\,\mathrm{ps}$ is comparable to the high-frequency
end of the segmental relaxation spectrum of polyurea and does not
permit full segmental relaxation between successive strain increments;
each strain step therefore represents a near-instantaneous deformation
followed by short-time relaxation, consistent with the effective strain
rates of order $10^{8}\,\mathrm{s}^{-1}$ reported in this work.}



{To suppress high-frequency thermal fluctuations while retaining physically relevant relaxation behavior, the resulting stress trajectories are post-processed using a Savitzky–Golay filter~\cite{savitzky1964smoothing} with a window length of 11 strain increments (corresponding to $\tau_{\mathrm{avg}} = 11 \,\mathrm{ps}$) and a quadratic polynomial. This choice removes sub-picosecond vibrational noise while preserving the segmental relaxation dynamics that govern viscoelastic dissipation in polyurea. The filter is zero-phase and locally polynomial, ensuring that peak stresses and relaxation curvature are not biased—an essential requirement for accurate evaluation of hysteresis. Sensitivity to the averaging procedure was assessed by varying the window length over ${7,11,21}$ increments and the polynomial orders ${2,3}$; the resulting variation in the stress response remained below $1\%$. Under this protocol, the total duration of each trajectory is $T_{\mathrm{traj}} = 1 \,\mathrm{ns}$, yielding effective strain rates of order $10^{8}\,\mathrm{s}^{-1}$, consistent with the intrinsic timescale limitations of fully atomistic polymer simulations~\cite{heyden2016all,manav2021polyurea_shock_md,dewapriya2021quantum}. These MD simulations resolve the segmental relaxation mechanisms (ps–ns) that dominate the viscoelastic response under the loading conditions considered.}

\begin{figure}[h!]
\centering
\subfigure[Bulk modulus]{\includegraphics[keepaspectratio=true,width=0.45\textwidth]{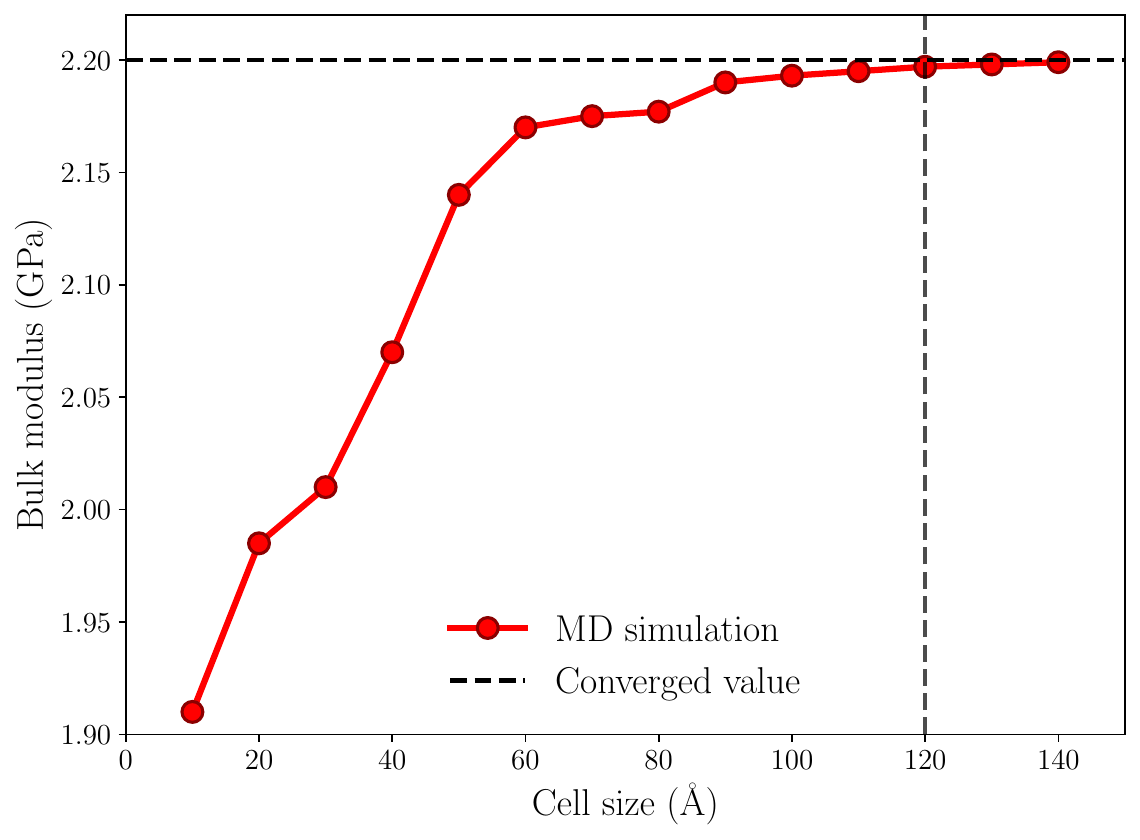}\label{Fig:PUBulk}}
\subfigure[stress strain plot]{\includegraphics[keepaspectratio=true,width=0.45\textwidth]{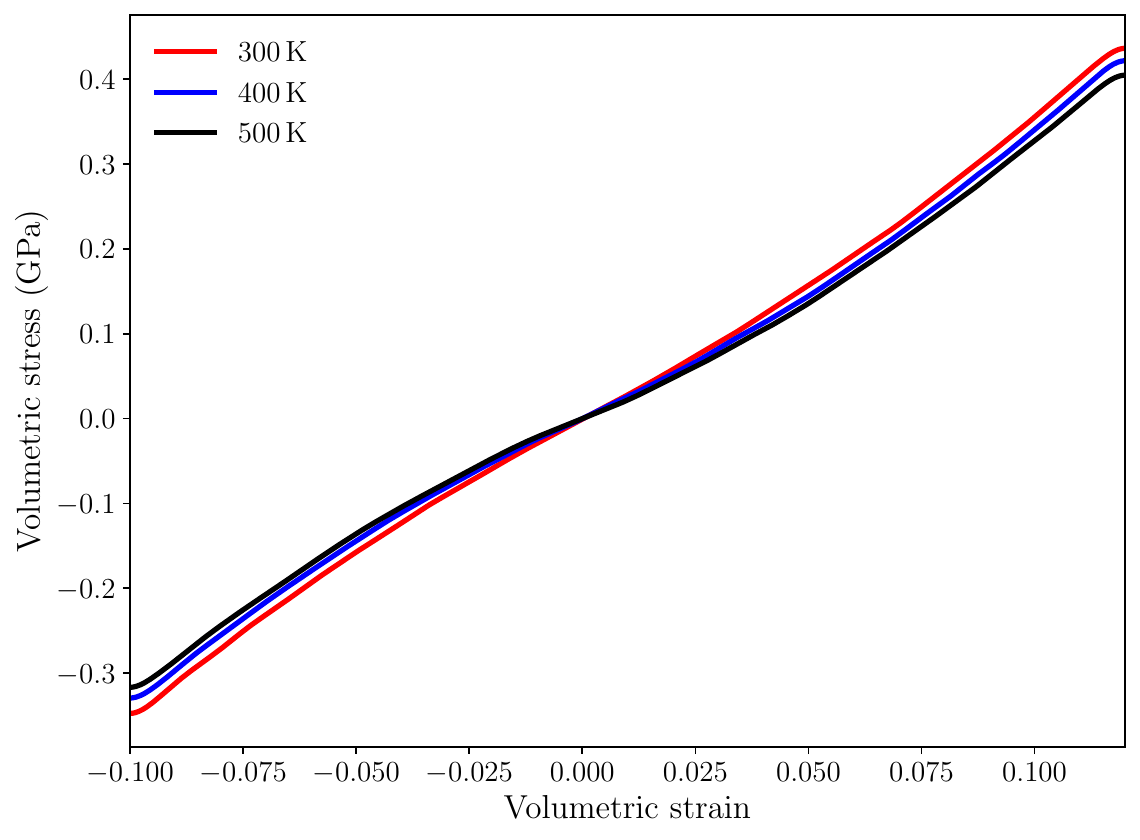}\label{Fig:PUStressStrain}}
{\caption{(a) shows the dependence of bulk modulus on the cell size of the RVE. (b) shows the dependence of stress on volumetric strain of polyurea at temperatures $300$, $400$, and $500$ K.}}
\end{figure}

To mitigate finite-size effects, the equilibrated cell was incrementally replicated, with each expansion followed by randomized rotation and translation of chains to disrupt artificial periodicity, and re-equilibrated under identical conditions. Figure \ref{Fig:PUBulk} shows the convergence of the bulk modulus with cell size, and we see that the bulk modulus converges at a cell size of $120$ \AA, which corresponds to a 525000 atoms supercell. The bulk modulus calculated using our atomistic simulation is $2.2$ GPa, which is in good agreement with previously reported atomistic values of $2.82$ GPa \cite{agarwal} from fully atomistic molecular dynamics simulations of polyurea as well as experimentally measured value of $2.57$ GPa at ambient conditions \cite{roland2007polyurea} 


Figure~\ref{Fig:PUStressStrain} shows the volumetric stress–strain response of polyurea obtained from atomistic simulations at temperatures of $300$~K, $400$~K, and $500$~K. The volumetric stress is $\sigma_{\mathrm{vol}} = \tfrac{1}{3}\mathrm{tr}(\boldsymbol{\sigma})$ and the volumetric strain is defined as $\varepsilon_{\mathrm{vol}} = \det\mathbf{F}-1$. From this figure, we see that the stress is influenced by temperature. The systematic decrease in slope with increasing temperature indicates enhanced compressibility due to increased free volume and thermal activation of molecular motion. 

\subsection{Recurrent Neural Operator atomistic surrogate}\label{subsec:AtomisticSurrogate}


To obtain the training data, we generated $1000$ strain paths at $300$ K and 500 strain paths each at $400$ K and $500$ K, with each trajectory discretized into $1000$ time steps. {Each strain path is generated using Equation \ref{Eq:deformationpath}, in which the independent components of the tensor $\nu_m$ are sampled from a Rademacher distribution taking values $\pm 1$. The resulting deformation gradient is then interpolated between knot points using cubic Hermite interpolation, yielding smooth, randomly oriented trajectories in the six-dimensional deformation-gradient space. The maximum magnitude $F_{\max}$ is $0.1$ to span the multiaxial deformation range encountered in the multiscale simulations described in Section~\ref{sec:multiscale}. The MD trajectories correspond to strain rates in the range $10^8$–$10^9~\mathrm{s^{-1}}$, consistent with the nanosecond timescales accessible to molecular dynamics. These rates lie within the regime used to calibrate the reference constitutive models employed for comparison and are representative of those encountered in the multiscale simulations considered here.} 

The RNO was first trained at $300$ K using the complete dataset, and the learned weights were subsequently used to initialize the models at $400$ K and $500$ K. During transfer learning, the kernel and decoder components of the neural operator were fine-tuned using the higher-temperature datasets, while leveraging the previously learned representation at 300 K. For each temperature, the dataset was partitioned into $80 \%$ training and $20 \%$ testing sets; accordingly, the $300$ K model used $800$ training and $200$ testing trajectories, while the $400$ K and $500$ K models each used $400$ training and $100$ testing trajectories for fine-tuning and evaluation. This transfer-learning strategy significantly reduces the amount of additional atomistic data required at elevated temperatures while maintaining predictive accuracy. Therefore, the total number of MD simulations to obtain the training data is $2000000$. 

\begin{figure}[h!]
\centering
\subfigure[Convergence plot for internal variables]
{\includegraphics[keepaspectratio=true,width=0.45\textwidth]{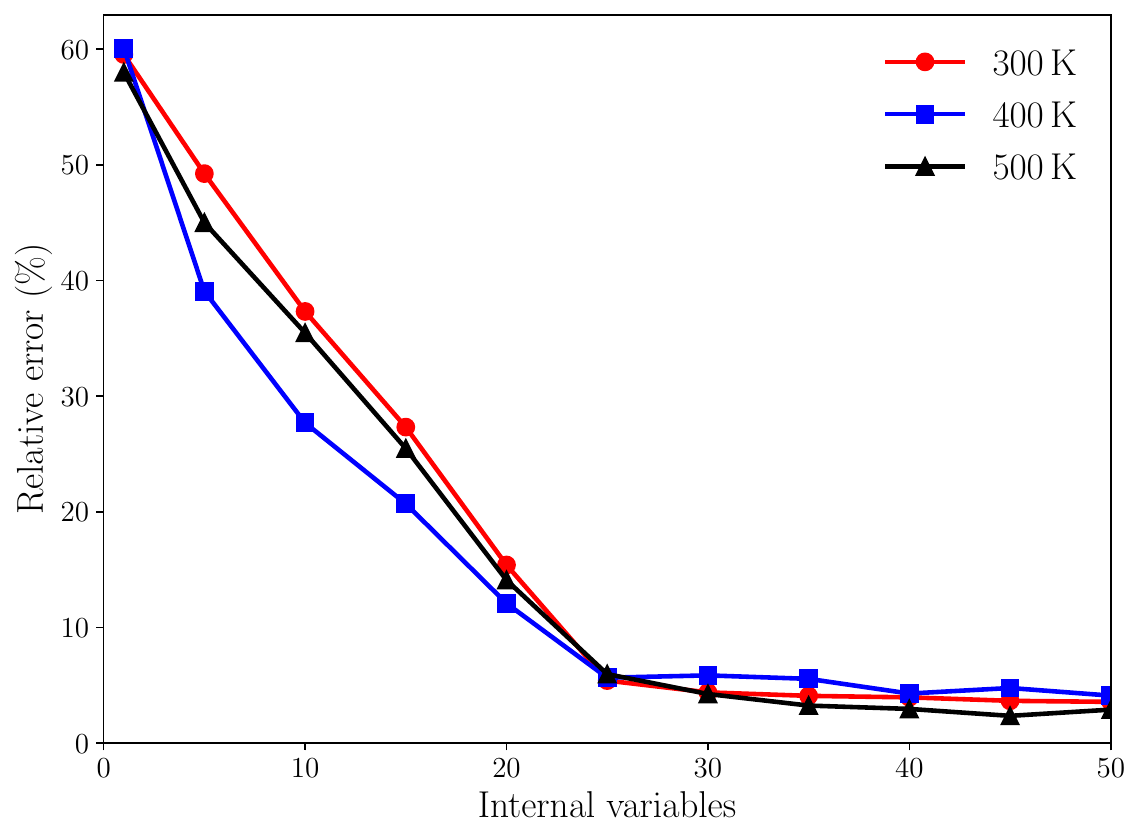}\label{Fig:intvar}}
\subfigure[relative error versus epoch]
{\includegraphics[keepaspectratio=true,width=0.45\textwidth]{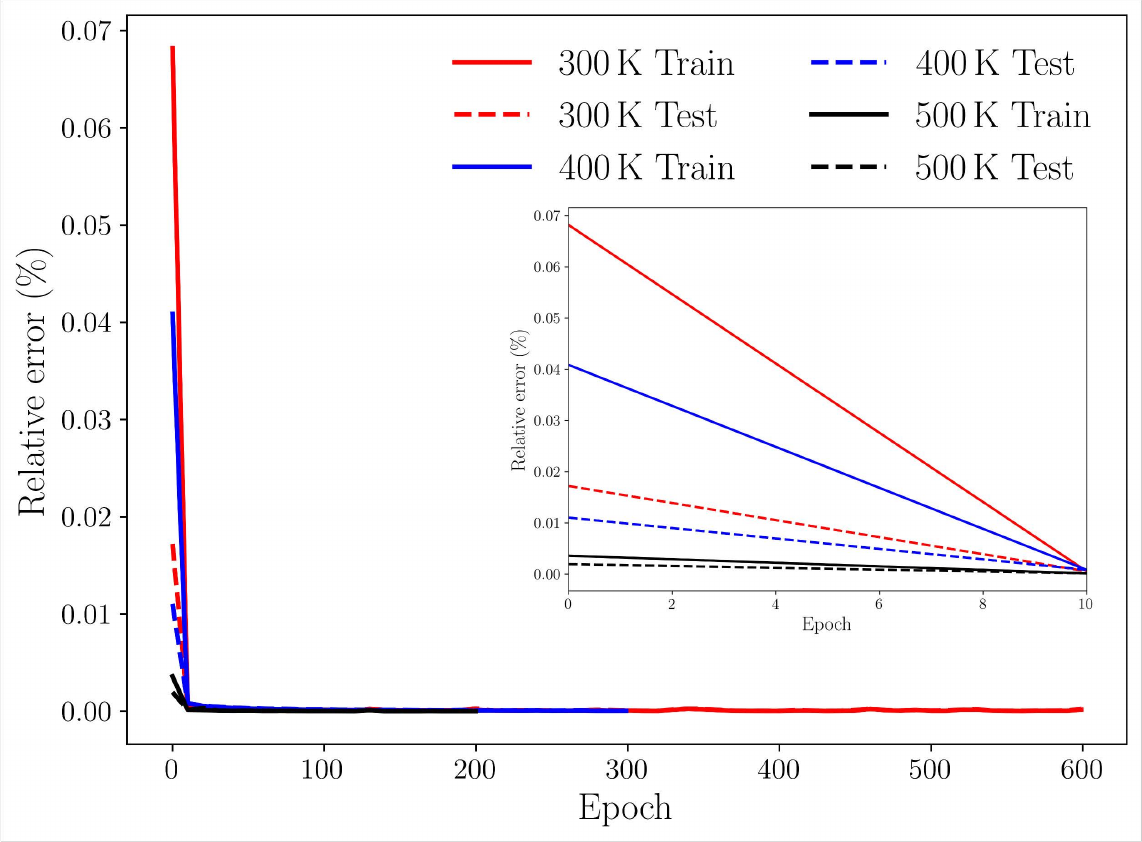}\label{Fig:Train/Testing error}\label{Fig:testingerror}}
{\caption{(a) Dependence of relative error on the number of internal variables for temperatures $300$ K, $400$ K, $500$ K. (b) shows the convergence of relative error over the number of epochs for temperatures $300$ K, $400$ K, $500$ K. Both training and testing errors are shown.}}
\end{figure}

Figure \ref{Fig:intvar} shows the dependence of the relative error on the number of internal variables 
$\xi$. The relative error decreases monotonically as $\xi$ increases from $1$ to $25$, after which it plateaus and remains nearly constant. Figure \ref{Fig:testingerror} shows the dependence of training and testing errors on the number of epochs. The rapid convergence observed in Figure \ref{Fig:testingerror}, particularly for the $400$ K and $500$ K cases, is a direct consequence of the transfer-learning strategy employed in this work. Specifically, the model trained at $300$ K provides a well-initialized representation of the underlying constitutive operator, capturing the dominant deformation-history-to-stress mapping and associated internal-variable dynamics. During transfer learning, the kernel and decoder components of the neural operator are fine-tuned using the higher-temperature datasets, allowing the model to adapt efficiently to temperature-dependent variations in material response. Since the fundamental viscoelastic mechanisms remain similar across temperatures, only a small number of epochs ($\sim 10$) are required for convergence at $400$ K and $500$ K. This significantly reduces the training cost while maintaining high predictive accuracy, as evidenced by the close agreement between training and testing errors across all temperatures.


\begin{figure}[h!]
\centering
{\includegraphics[keepaspectratio=true,width=1.0\textwidth]{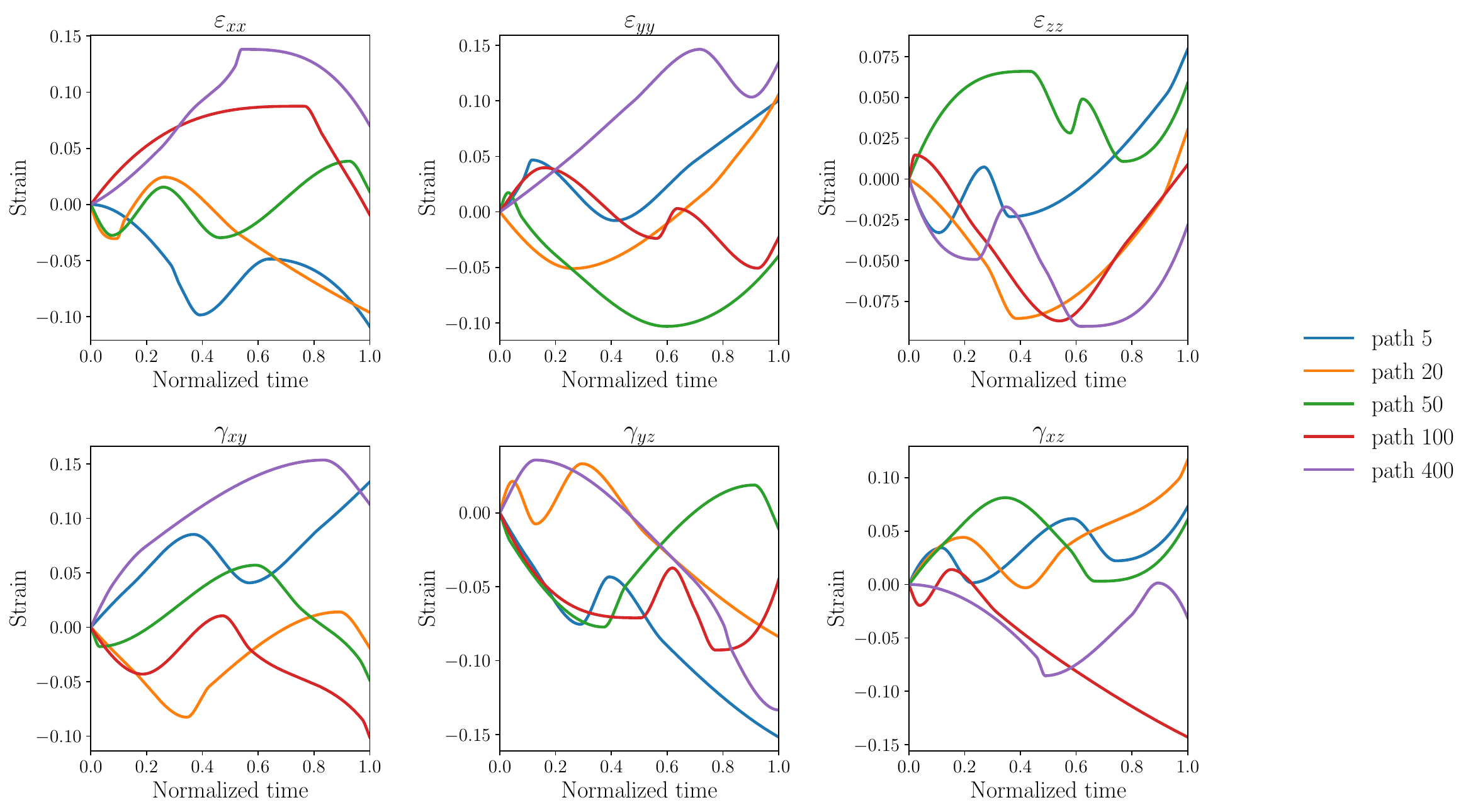}\label{Fig:strain}}
{\caption{Components of strain tensor shown for five trajectories. The normal strains are shown in the top panel, and shear strains are shown in the bottom panel. {Normalized time is defined as $\hat{t} = t / T_{\mathrm{total}}$, with $T_{\mathrm{total}} = 1\,\mathrm{ns}$ is the total time for each MD simulation.}\label{Fig:strain}}}
\end{figure}

\begin{figure}[h!]
\centering
{\includegraphics[keepaspectratio=true,width=1.0\textwidth]{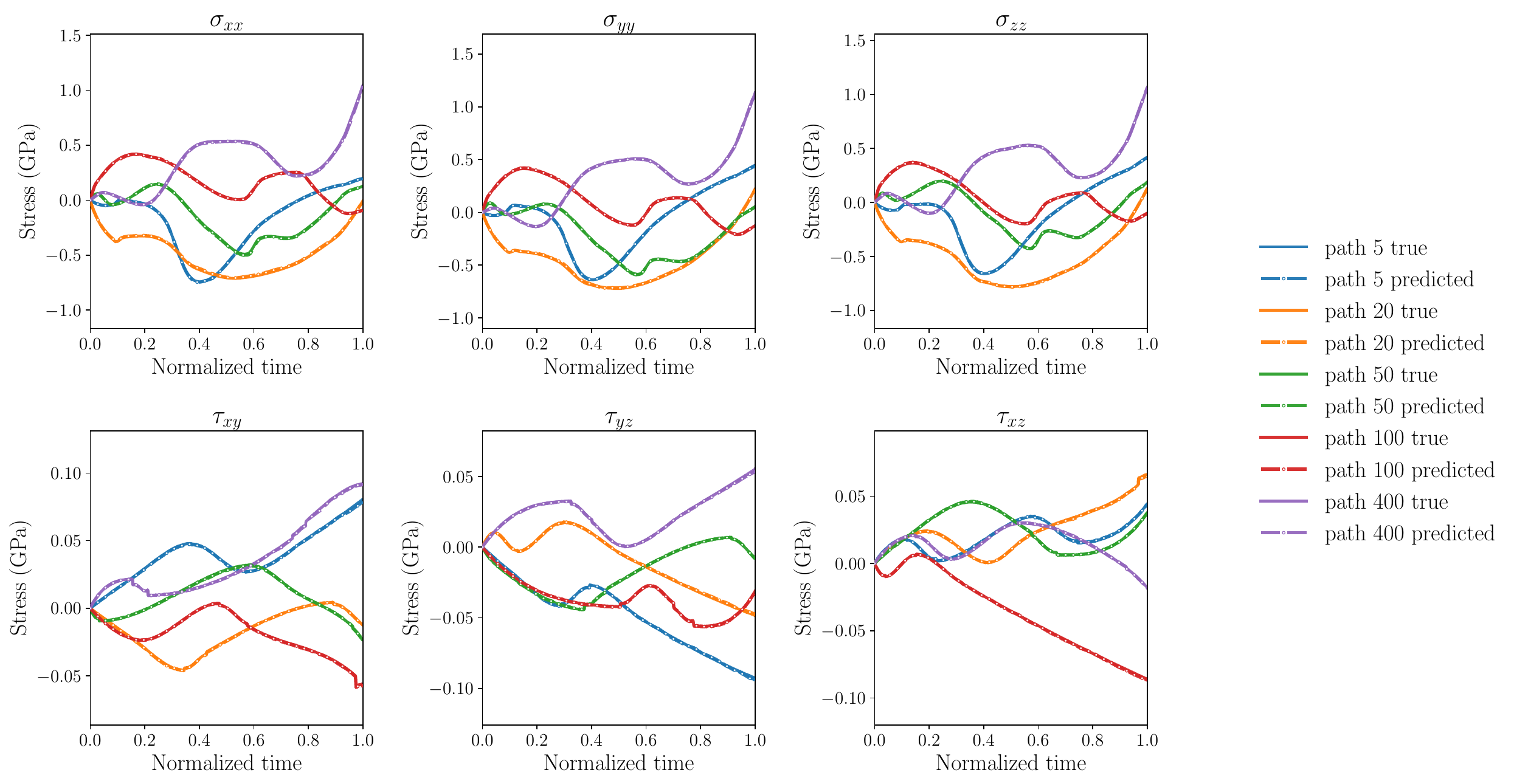}}
{\caption{Stress prediction and convergence for RNO at 300K.  The uniaxial stresses are shown in the top panel, and shear stresses are shown in the bottom panel.}\label{Fig:stress_300}}
\end{figure}


Having established the convergence of error with internal variables and the efficacy of transfer learning across temperatures, we evaluate the response calculated by the surrogate. For this, we use $25$ internal variables. In Figure \ref{Fig:strain}, we plot the components of the strain tensors for five representative strain paths used in our work. These strain paths are taken from the test set. We plot the true stress calculated from atomistic simulations as well as the stress predicted by the atomistic surrogate in Figure \ref{Fig:stress_300} for $300$ K. The true and the predicted values of the stresses for $400$ K and $500$ K are shown in Appendix A. It is clear from these plots that the predicted values of stress calculated using the atomistic surrogate are in excellent agreement with the true values of stress calculated using molecular simulations, therefore accurately capturing both the magnitude and temporal evolution of the stress response. Furthermore, the surrogate captures important viscoelastic features of polyurea, including rate-dependent stress buildup, stress relaxation, and strain softening under complex multiaxial loading.  In particular, the phase lag between stress and strain, as well as the gradual decay of stress following loading reversals, are consistently reproduced by the atomistic surrogate across all stress components,  demonstrating that the learned internal variables encode coupled memory effects rather than isolated component-wise behavior.
{Throughout this section, we employ a normalized time $\hat{t} = t / T_{\mathrm{total}}$, where $T_{\mathrm{total}} = 1\,\mathrm{ns}$ denotes the duration of each MD training trajectory.}

Overall, the atomistic surrogate generalizes reliably to previously unseen strain histories and this fidelity is maintained across the full temperature range considered, indicating that the surrogate captures the temperature-dependent viscoelastic response of polyurea. Therefore, the atomistic surrogate can provide an accurate, computationally efficient replacement for direct fine-scale simulations for history-dependent constitutive modeling.


\section{Multiscale simulations of viscoelastic composites}\label{sec:multiscale}
We perform multiscale simulations of polyurea composites using the atomistic surrogate discussed in the previous section \ref{Sec:Learning}. {The atomistic surrogate is benchmarked against two constitutive models whose parameters are taken directly from experimental literature and are independent of our MD data. These are (a) the pressure--shear plate-impact-derived quasi-linear viscoelastic model of Clifton, Wang and Jiao \cite{clifton2016physically}, originally calibrated against plate-impact experiments at $\sim 18$~GPa and strain rates of $10^{5}$--$10^{6}\,\mathrm{s}^{-1}$; and (b)~the Johnson--Cook model~\cite{johnson1983constitutive} with the parameter set of Choi and Chahine~\cite{choi2015polyurea}, obtained by fitting Split Hopkinson Pressure Bar data for polyurea at rates up to $1.2 \times 10^{4}\,\mathrm{s}^{-1}$.  Because these parameter sets were calibrated using experiments that are entirely independent of the MD trajectories used to train the RNO, the comparison serves as a validation of the RNO surrogate rather than a self-consistency check.} The details of these material models and the parameters are discussed in Appendix B.

We perform three tests. First is a cyclic loading of a three-dimensional specimen. Second is the Taylor impact test, where a three-dimensional cylindrical sample impacts a wall. Finally, the plate impact test where a three-dimensional polyurea plate is impacted by an rigid impactor. We use ABAQUS \cite{abaqus2016theory} for all the simulations. The RNO atomistic surrogate and the Clifton material models are implemented through VUMAT subroutines.

{The RNO surrogate returns the stress and updates the latent internal state at each quadrature point, and it does not predict thermal transport. In all three continuum examples, the temperature field is instead evolved using ABAQUS/Explicit's coupled thermo-mechanical integration scheme. The thermal response is driven
by externally prescribed material parameters that are identical across the constitutive models considered here, namely a heat source arising
from local mechanical dissipation and Fourier heat conduction. The heat source is defined as the difference between the stress power returned by the constitutive routine and the rate of recoverable
elastic strain energy, scaled by the Taylor--Quinney coefficient $\beta_{\mathrm{TQ}} = 0.9$~\cite{taylor1934latent,rosakis2000thermodynamic}. Heat conduction is governed by thermal conductivity $k = 0.19~\mathrm{W\,m^{-1}\,K^{-1}}$, specific heat
$c_p = 1500~\mathrm{J\,kg^{-1}\,K^{-1}}$, and mass density
$\rho = 1120~\mathrm{kg\,m^{-3}}$, representing solid polyurea and
related solid polyurethane elastomers~\cite{xie2024pu_thin_films,vankrevelen2009properties};the density is consistent with the equilibrated MD supercell of
Sec.~\ref{subsec:AtomisticSimulations}. The corresponding thermal diffusivity is
$\alpha = k/(\rho c_p) \approx 1.1 \times 10^{-7}~\mathrm{m^{2}\,s^{-1}}$,
which gives a thermal diffusion length
$\sqrt{\alpha\,\tau} \lesssim 5~\mu\mathrm{m}$ for the longest dynamic
event considered here ($\tau = 300~\mu\mathrm{s}$ for the plate impact test). This is several orders of magnitude smaller than the specimen dimensions ($\geq 5~\mathrm{mm}$), so heat transport over the microsecond-to-hundreds-of-microseconds timescales of the simulations is conduction-limited and the temperature evolution is dominated by
local mechanical dissipation. The reported temperature fields are therefore only weakly sensitive to moderate variations in $k$, $c_p$,and $\rho$ within experimentally reported ranges under the present loading conditions. These quantities are external inputs to the FEM solver and are independent of the RNO surrogate.}

\subsection{Cyclic loading}

\begin{figure}[h!]
\centering
 {\includegraphics[keepaspectratio=true,width=0.7\textwidth]{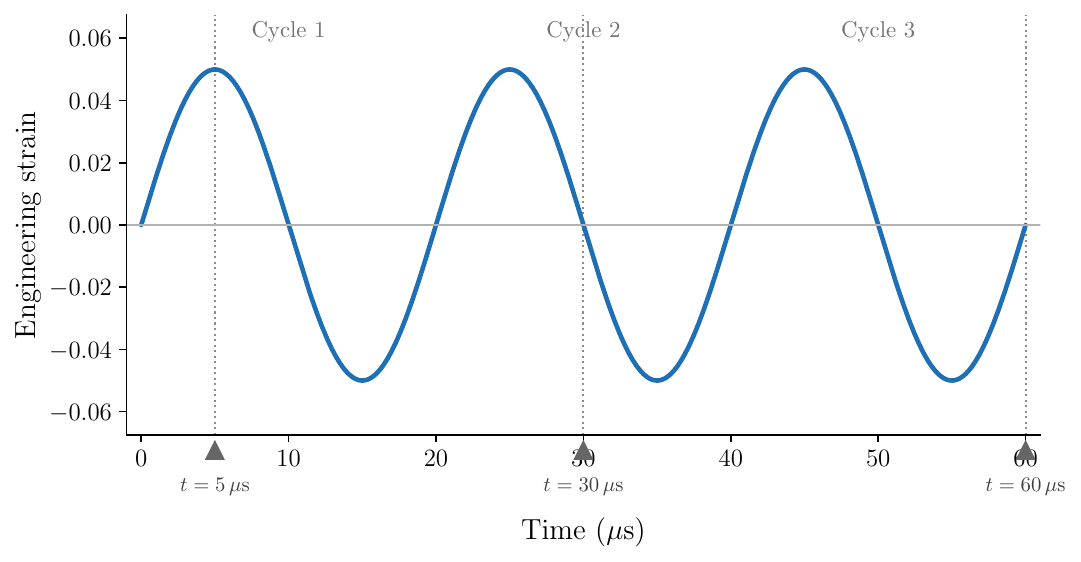}}
   {\caption{Applied nominal engineering strain $\varepsilon(t)$ during the cyclic-loading test. Three sinusoidal cycles are imposed over $0\le t \le 60\,\mu\mathrm{s}$ with period $T=20\,\mu\mathrm{s}$ and peak amplitude $\pm 0.05\%$, smoothed by a $2\,\mu\mathrm{s}$ cosine ramp at $t=0$.}
    \label{Fig:DogboneCyclic_loading}}
\end{figure}

\begin{figure}[h!]
\centering
{\includegraphics[keepaspectratio=true,width=1.0\textwidth]{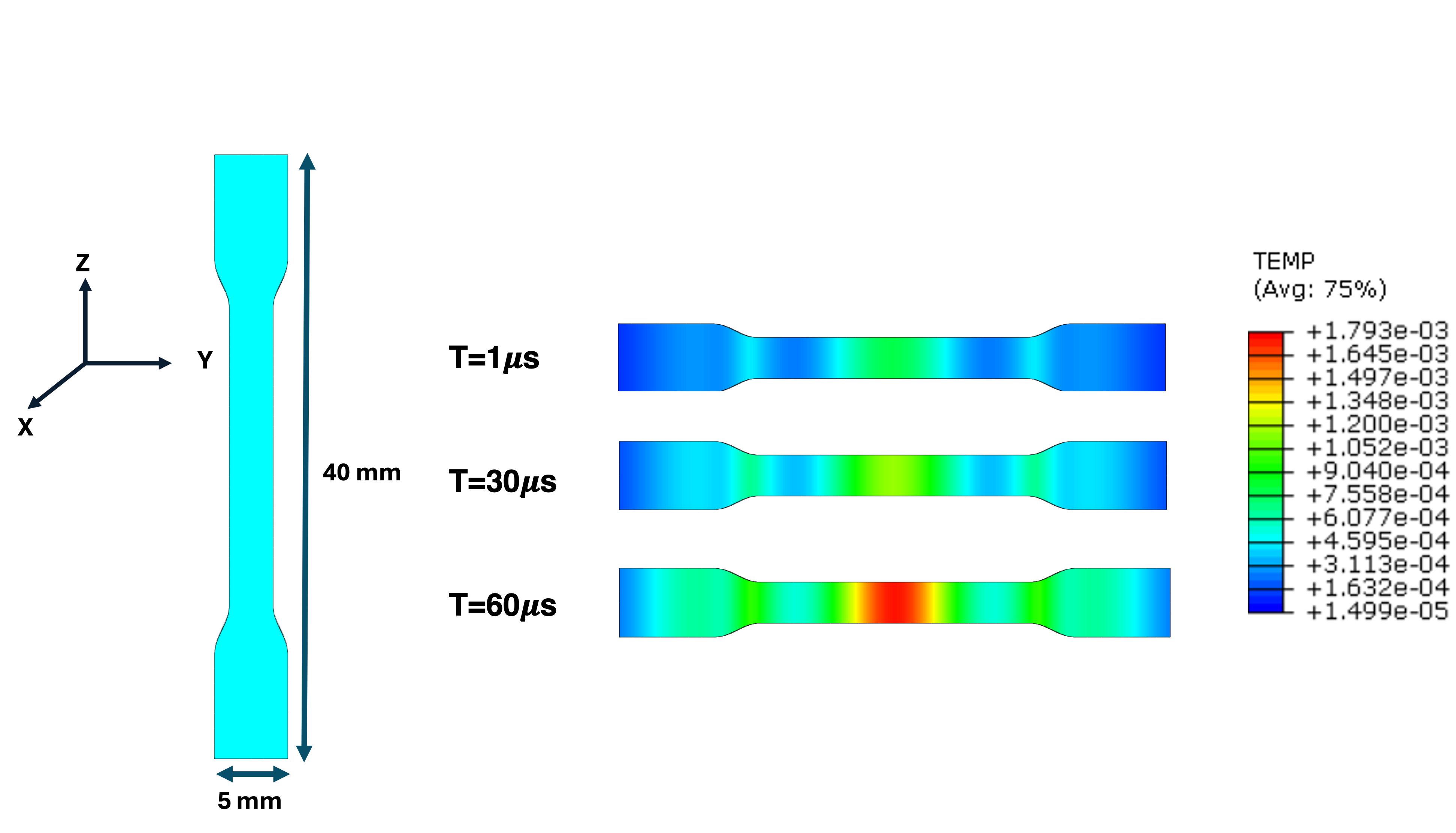}}
{\caption{The geometry of the {dogbone} specimen is shown on the left. {The temperature change from 300 K, calculated at different times during the simulation is shown on the right. The unit of temperature is Kelvin.}}\label{Fig:DogboneTemp}}
\end{figure}



We apply cyclic loading, as shown in Figure \ref{Fig:DogboneCyclic_loading}, to a three-dimensional polyurea specimen with dimensions of 40 × 5 × 1 mm in the axial, transverse, and thickness directions, respectively. In Figure \ref{Fig:DogboneTemp}, the geometry of the specimen is shown. The end sections are enlarged to facilitate the application of boundary constraints, while the central reduced-width region (gauge section) promotes deformation localization away from the boundaries. The specimen is modeled as a three-dimensional solid and discretized using C3D8RT reduced-integration hexahedral elements \cite{abaqus2016theory,belytschko2013nonlinear}. These elements are well suited for large-deformation explicit dynamics and provide efficient, stable per-integration-point evaluation of constitutive models. Reduced integration alleviates volumetric locking in nearly incompressible materials and significantly reduces computational cost, which is important when using history-dependent material models~\cite{hughes2012finite}. The mesh contains {672 elements and 1197 nodes}. 

The loading is applied along the axial direction (z-axis) using distributing coupling constraints that tie each end surface to a reference point. This kinematic constraint enforces uniform displacement of the end surfaces while allowing internal stress redistribution, thereby mimicking rigid grips without over-constraining the specimen. The bottom end surface is fixed to eliminate rigid-body motion, while cyclic displacement is prescribed at the top end surface. The imposed displacement amplitude corresponds to 0.05--0.10\% nominal engineering strain and is applied as a smooth periodic function within a single explicit dynamic step.

The material response is modeled using (a) the RNO-based atomistic surrogate, (b) the physics-based viscoelastic polyurea model of Clifton, and (c) the Johnson–Cook model. For comparison, all geometric, boundary, and loading parameters are held fixed across simulations.

Figure \ref{Fig:DogboneTemp}  shows the spatial distribution of the increase in temperature during cyclic loading at times $1 \mu$s, $30 \mu$s, and $60 \mu$s when the RNO surrogate is used. With time, temperature increases and spreads along the gauge length, indicating cumulative dissipation of mechanical work. The absence of spurious temperature spikes or numerical oscillations indicates stable thermomechanical coupling and consistent energy balance in the surrogate model.

The observed temperature patterns are consistent with viscoelastic energy dissipation, where part of the mechanical work is converted into heat during cyclic loading ~\cite{lakes2009viscoelastic, ward2012mechanical}. In polymers, repeated strain reversals activate molecular segmental motion and internal friction, leading to localized heat generation in regions experiencing cyclic deformation.

\begin{figure}[h!]
\centering
{\includegraphics[keepaspectratio=true,width=1.0\textwidth]{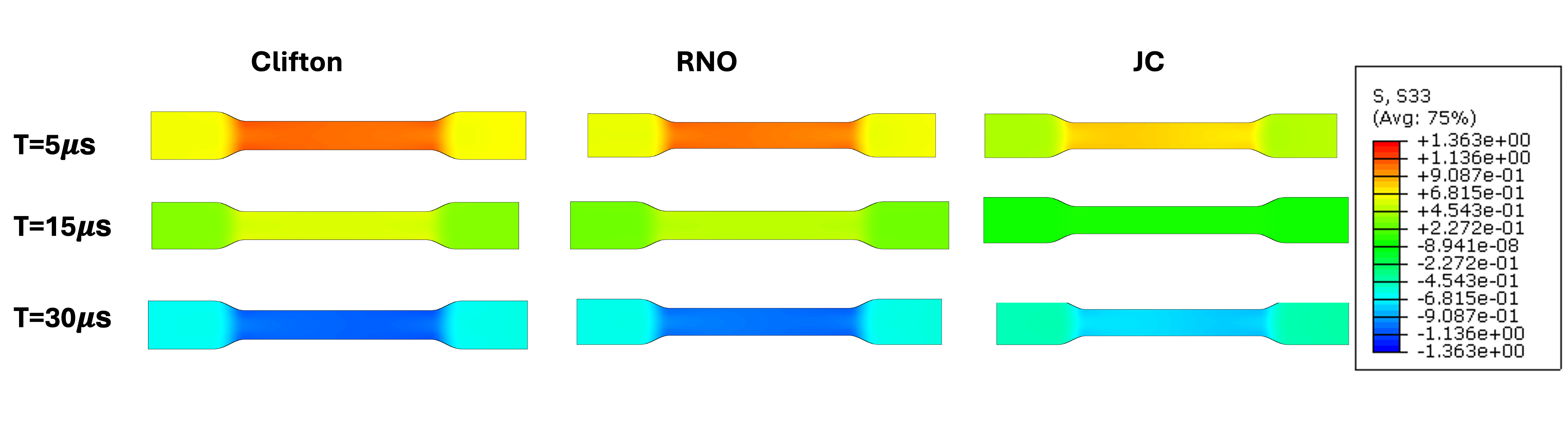}}
{\caption{Comparison of the uniaxial stress {(MPa)} in the $33$ direction for the Clifton, RNO atomistic surrogate and Johnson-Cook material models at three different times. \label{Fig:DogboneStress}}
}
\end{figure}

{To validate, we derive an analytical estimate of the temperature rise. For a sinusoidal strain, the loss modulus is $E\,\sin{\delta}$, where $\delta$ is the loss angle. The energy dissipated per unit volume in one cycle is $W=\pi \varepsilon_{max}^2\,E\,\sin{\delta}$. Over $N$ cycles, a fraction $\beta_{\mathrm{TQ}}$ of the dissipated work is converted to heat. Therefore, the heat generated per unit volume is $Q=\beta_{\mathrm{TQ}}\,N\,\pi \varepsilon_{max}^2\,E\,\sin{\delta}$. Assuming adiabatic heat generation and neglecting conduction over the loading event, $Q=\rho\,c_p\,\Delta \theta$. Hence, we obtain $\Delta \theta=(\beta_{\mathrm{TQ}}\,N\,\pi \varepsilon_{max}^2\,E\,\sin{\delta})/(\rho\,c_p)$. Using $\beta_{\mathrm{TQ}}=0.9$ ~\cite{taylor1934latent,rosakis2000thermodynamic}, $N=3$, $\varepsilon_{max}=5\times10^{-4}$, $E=2.82$ GPa \cite{agarwal}, $\delta=0.46$ rad \cite{jia2016experimentally}, $\rho=1120$ kg m$^{-3}$ ~\cite{vankrevelen2009properties}, $c_p=1500$ J\,kg$^{-1}$\,K$^{-1}$ ~\cite{xie2024pu_thin_films}, we obtain the maximum rise in temperature $\Delta \theta$ as $1.6$ mK, which is in excellent agreement with the increase in temperature calculated using RNO atomistic surrogate of $1.8$ mK. }


Figure \ref{Fig:DogboneStress} shows the axial stress ($S_{33}$) distribution in the specimen for the Clifton, RNO atomistic surrogate, and Johnson–Cook constitutive models at three different times ($5\mu$s, $15\mu$s, and $30\mu$s) during cyclic loading. {The cyclic displacement is applied from rest, with an identical prescribed loading history (amplitude and frequency) imposed on all three models, ensuring synchronous kinematic input. However, the responses are not phase-synchronous, as each model exhibits a distinct relaxation time and hence a different phase lag relative to the imposed strain. Consequently, their instantaneous stress fields do not coincide at a given wall-clock time.} {The Clifton model and the RNO surrogate yield nearly identical stress fields, exhibiting the same spatial distribution and peak stress magnitudes that differ by less than $\sim 10\%$. This agreement demonstrates that the RNO surrogate accurately reproduces the salient features of the Clifton model, despite being trained exclusively on MD-generated trajectories. In contrast, the peak stresses predicted by the Johnson--Cook model are approximately $28\%$ lower than those obtained with the Clifton model. This discrepancy arises because the Johnson--Cook formulation is governed by the elastic modulus of polyurea, whereas the viscoelastic models are controlled by the high-frequency complex modulus. Nevertheless, all three models predict a tensile stress concentration within the gauge section and lower stress levels in the flange region, indicating that the overall stress distribution is primarily dictated by the dogbone specimen geometry.} {We also note that the peak stress calculated using the RNO atomistic surrogate is $1.30$ MPa, which is in good agreement with the analytical estimate of $E\,\varepsilon_{max}$ = $2.82$ GPa $\times$ $5 \times 10^{-4}$ = $1.41$ MPa}

\begin{figure}[h!]
\centering
{\includegraphics[keepaspectratio=true,width=0.7\textwidth]{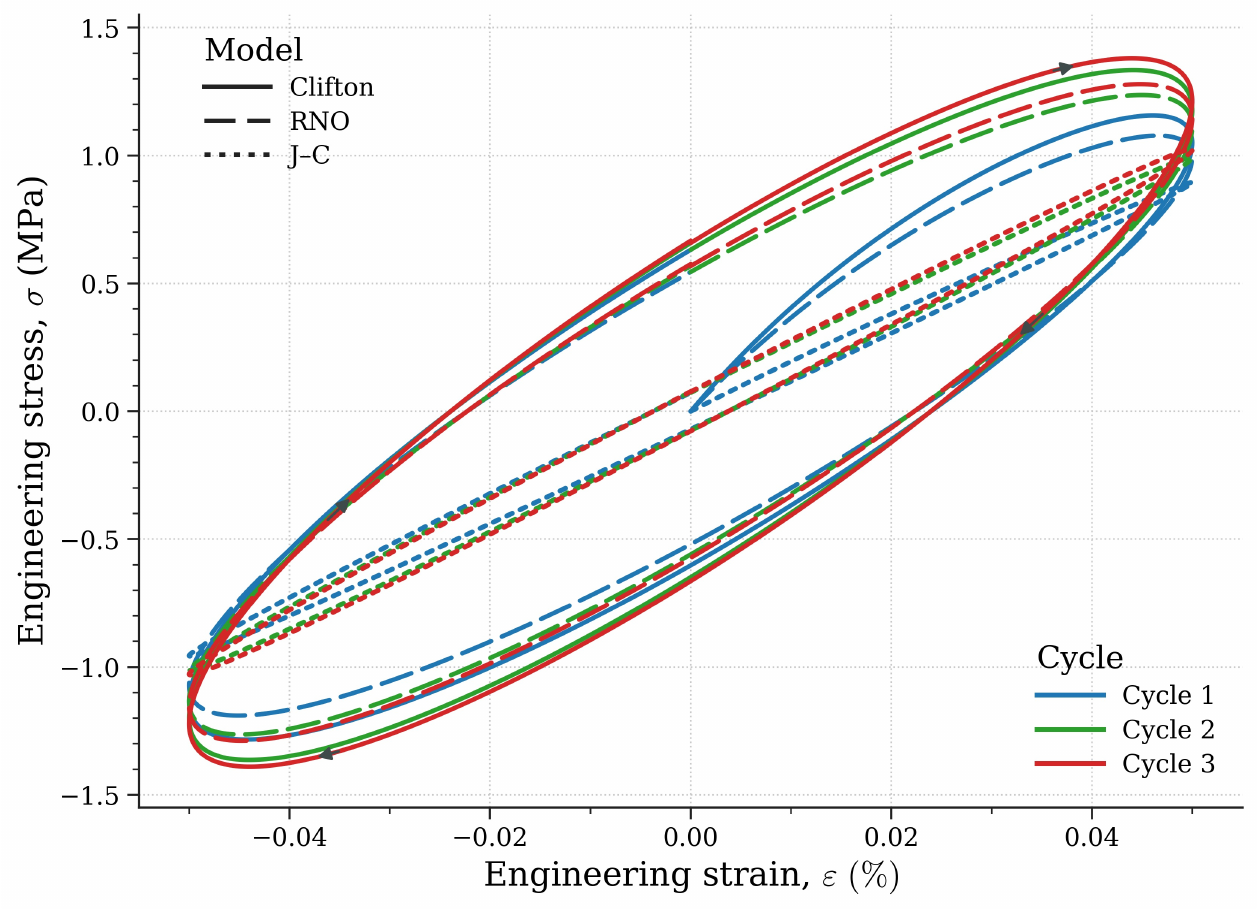}}
{\caption{{Engineering stress--strain hysteresis loops for three consecutive loading cycles, comparing Clifton (solid line), the RNO atomistic surrogate (dashed line), and Johnson--Cook (dotted line). Black arrows indicate clockwise traversal direction, common to all loops.}}\label{Fig:DogboneCyclic}}
\end{figure}

In Figure \ref{Fig:DogboneCyclic} we plot the engineering stress–strain hysteresis loops for three consecutive loading cycles. It is evident from this plot that Clifton's model exhibits noticeable hysteresis with smooth loading–unloading transitions, effectively capturing viscoelastic dissipation and stress relaxation behavior characteristic of polyurea. Also seen from the Figure, the RNO atomistic surrogate is in close agreement with the hysteresis behavior calculated using Clifton's model across all cycles. Both the slope of the loading and unloading branches and the enclosed area of the hysteresis loop are in excellent agreement, demonstrating that the neural-operator surrogate captures the correct phase lag between stress and strain as well as the cumulative dissipation associated with cyclic deformation of history-dependent materials \cite{roland2007polyurea}. On the other hand, the Johnson–Cook model exhibits narrower hysteresis loops with comparatively less energy dissipation. Stress–strain paths are closer to elastic–plastic behavior \cite{johnson1983constitutive,belytschko2013nonlinear}, with limited relaxation during unloading and a smaller area of the hysteresis loop, consistent with its metal-plasticity assumptions and lack of intrinsic material memory \cite{johnson1983constitutive,simo1998computational}. {The Johnson–Cook loops remain thermodynamically consistent, with irreversible plastic dissipation converted into heat through the Taylor--Quinney factor, $\beta_{\mathrm{TQ}} = 0.9$~\cite{taylor1934latent}. The narrower loop simply reflects the absence of intrinsic viscoelastic dissipation in the Johnson–Cook formulation rather than any violation of the second law~\cite{simo1998computational}.}

To summarize, the cyclic loading test results demonstrate that the RNO atomistic surrogate faithfully reproduces the history-dependent, dissipative response of polyurea under repeated loading. Agreement with the experimentally derived Clifton's material model is observed at both the field level (stress and temperature distributions) as well as the global level (stress–strain hysteresis). In contrast, the Johnson–Cook model does not capture essential features of cyclic loading of polyurea.

\subsection{Taylor impact}
\begin{figure}[h!]
\centering
{\includegraphics[keepaspectratio=true,width=1.0\textwidth]{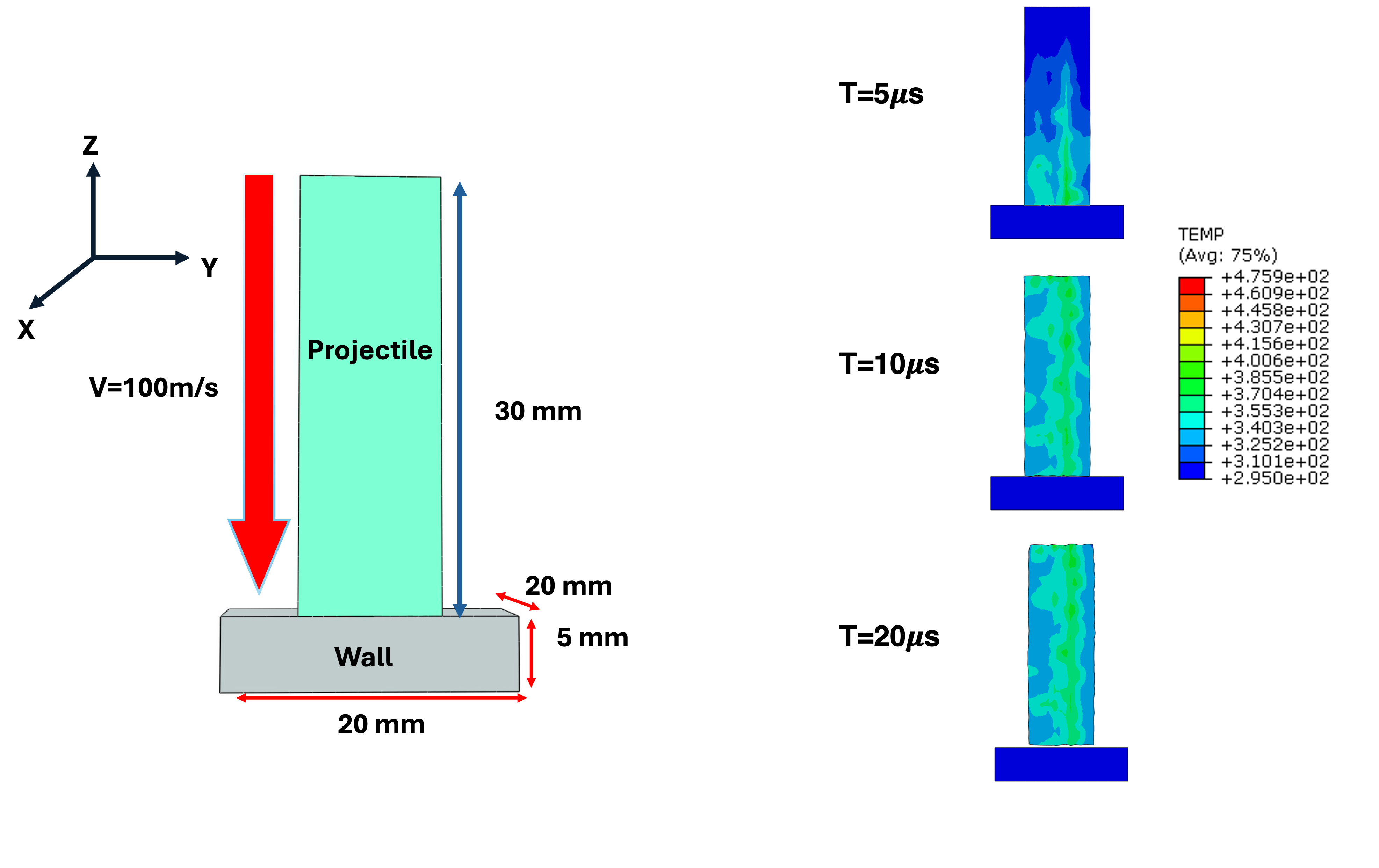}}
{\caption{The geometry of the Taylor impact test are schematically shown on the left. The evolution of temperature {(K)} across the midplane of the cylinder during the simulation is shown on the right.}\label{Fig:TaylorTemp}}
\end{figure}

We perform the Taylor impact test where a polyurea projectile of height 30 mm and diameter $10$ mm, traveling with an initial velocity $V = 100$ m/s, impacts a rigid frictionless wall of dimensions $20 \times 20 \times 5$ mm at time $t = 0$ s. This is schematically shown in Figure \ref{Fig:TaylorTemp}. The selected geometry promotes predominantly uniaxial compressive deformation while suppressing bending and boundary-induced artifacts, consistent with classical Taylor impact assumptions \cite{taylor1948impact,sarva2007polycarbonate}. The wall is modeled as a rigid elastic solid with Young’s modulus $E = 210$ GPa and is fully constrained in all translational degrees of freedom, thereby acting as an effectively rigid support throughout the simulation. 

Contact between the projectile and the wall is enforced using hard normal contact with frictionless tangential behavior in order to isolate compressive impact physics without introducing additional shear dissipation. The computational domain is discretized using C3D8RT reduced-integration brick elements \cite{abaqus2016theory,belytschko2013nonlinear}. The projectile–wall assembly consists of approximately $3,928$ elements and $4,848$ nodes, resulting in one constitutive evaluation per element per time increment. This discretization provides a robust and computationally efficient representation for large-deformation explicit dynamics, while enabling local per-integration-point evaluation of both Clifton's and RNO atomistic surrogate constitutive models.

Figure \ref{Fig:TaylorTemp} also shows the evolution of the temperature field as calculated by the RNO atomistic surrogate constitutive model during the Taylor impact test. The contours depict the spatial distribution of temperature within the polyurea specimen at representative times corresponding to early, intermediate, and late stages of the impact simulation. At early stage ($t=5\mu$s), temperature rise is localized near the impact interface, coinciding with regions of high compressive stress and strain rate. This localized heating reflects the conversion of mechanical work into thermal energy during the initial wave-dominated loading phase. As the impact progresses ($t=10\mu$s), the temperature field extends axially along the specimen, forming a continuous band aligned with the loading direction. The temperature distribution becomes smoother and more spatially distributed, indicating the spreading of deformation and dissipation beyond the immediate contact region. Peak temperatures remain concentrated along the centerline, consistent with the dominant uniaxial compression and associated inelastic deformation. At later times ($t=20\mu$s), the temperature field exhibits a stabilized spatial pattern with reduced gradients, suggesting that the majority of impact-induced heating has occurred and that the system has transitioned into a post-impact relaxation regime. Residual temperature elevations persist along the deformation path, reflecting the cumulative effect of dissipative processes during the impact. 

Overall, the temperature contours calculated using the RNO atomistic surrogate constitutive model indicate a physically consistent evolution of thermomechanical coupling, with localized heating during early high-rate deformation followed by redistribution and stabilization as the impact event concludes.


\begin{figure}[h!]
\centering
{\includegraphics[keepaspectratio=true,width=1.0\textwidth]{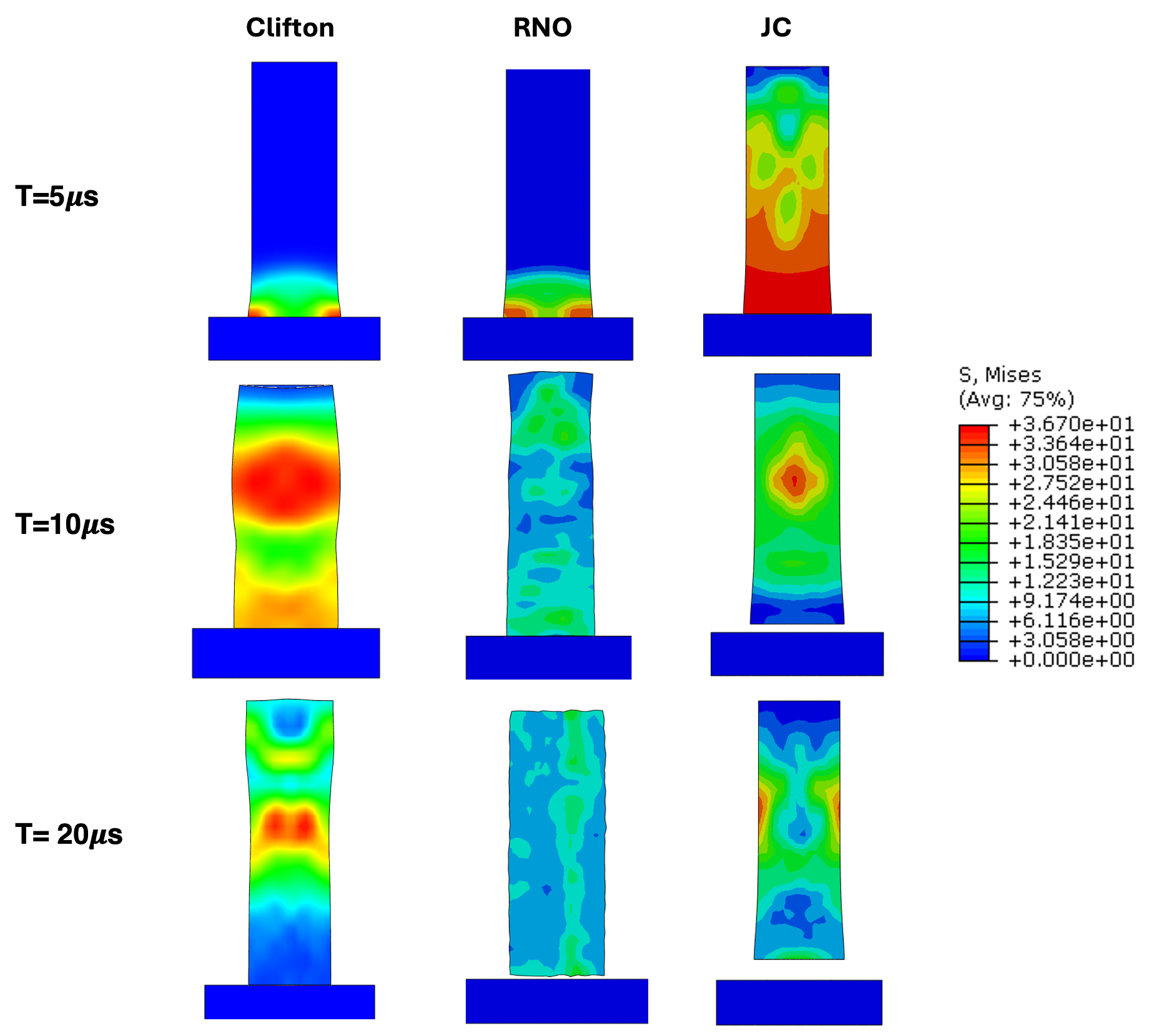}}
{\caption{Comparison of von Mises stress {(MPa)} for the three material models is shown at three different times for the Taylor impact test}\label{Fig:TaylorStress}}
\end{figure}

In Figure \ref{Fig:TaylorStress}, we compare the evolution of the von Mises stress for the Taylor impact test using Clifton’s constitutive model, the RNO atomistic surrogate, and the Johnson--Cook constitutive model. The von Mises stress is defined as
\[
\sigma_{\mathrm{vm}} \;=\; \sqrt{\frac{3}{2}\,\boldsymbol{s}:\boldsymbol{s}}
\;=\;
\sqrt{\frac{1}{2}\left[(\sigma_{11}-\sigma_{22})^2+(\sigma_{22}-\sigma_{33})^2+(\sigma_{33}-\sigma_{11})^2\right]
+3\left(\sigma_{12}^2+\sigma_{23}^2+\sigma_{31}^2\right)},
\]
where $\boldsymbol{s}=\boldsymbol{\sigma}-\tfrac{1}{3}\mathrm{tr}(\boldsymbol{\sigma})\boldsymbol{I}$ is the
deviatoric Cauchy part of the stress tensor. The stress contours are shown at $t=5$, $10$, and $20~\mu$s, corresponding to early impact, peak deformation, and post-impact relaxation regimes, respectively. All constitutive models are evaluated using identical geometry, boundary conditions, contact definitions, and mesh discretizations. At early time ($t=5~\mu$s), the response is dominated by compressive wave propagation initiated at the projectile--anvil interface. High stress concentrations develop near the contact region and propagate axially into
the specimen, forming a well-defined stress front characteristic of high-rate, predominantly uniaxial compression. We see the RNO atomistic surrogate to be in good agreement with Clifton's constitutive model. At intermediate time ($t=10 \mu$ s), corresponding to the peak deformation regime, the stress field extends over a larger portion of the specimen length. Peak stresses are redistributed away from the immediate contact region, and the stress contours indicate the development of a nonuniform axial stress profile associated with significant inelastic deformation and evolving material response under sustained loading. The maximum stress obtained using the RNO atomistic surrogate is smaller compared to Clifton's constitutive model or the Johnson-Cook model. By late time ($t=20 \mu$ s), the stress field transitions into a relaxation-dominated regime. Peak stress magnitudes decrease, and stresses become more broadly distributed along the specimen. 

{The contours in Figures \ref{Fig:TaylorTemp} and \ref{Fig:TaylorStress} exhibit minor departures from exact axisymmetry, which are most visible in the higher-resolution RNO-surrogate results. These deviations are numerical in origin and arise from two sources. First, the underlying three-dimensional hexahedral mesh is not perfectly axisymmetric; elements on the cylindrical boundary necessarily break continuous rotational symmetry, and small azimuthal variations in element orientation and contact discretization can be amplified in the rendered contour fields. Second, the RNO acts as a pointwise constitutive operator and therefore yields slightly different responses for elements with nearly identical deformation histories; these small discrepancies can be further amplified by contact interactions and wave reflections. The same mesh-induced asymmetry is present in the Clifton and Johnson–Cook solutions, but is less apparent due to their constitutive responses being less sensitive to small element-to-element variations in deformation history, which tends to suppress the visual amplification of numerical discretization effects in the contour fields. Importantly, the constitutive updates themselves are formulated in a pointwise manner and preserve rotational invariance at the material level; the observed asymmetry therefore reflects the spatial discretization and numerical propagation effects of the three-dimensional finite element model rather than any intrinsic artifact of the RNO surrogate.}

\begin{figure}[h!]
\centering
{\includegraphics[keepaspectratio=true,width=1.0\textwidth]{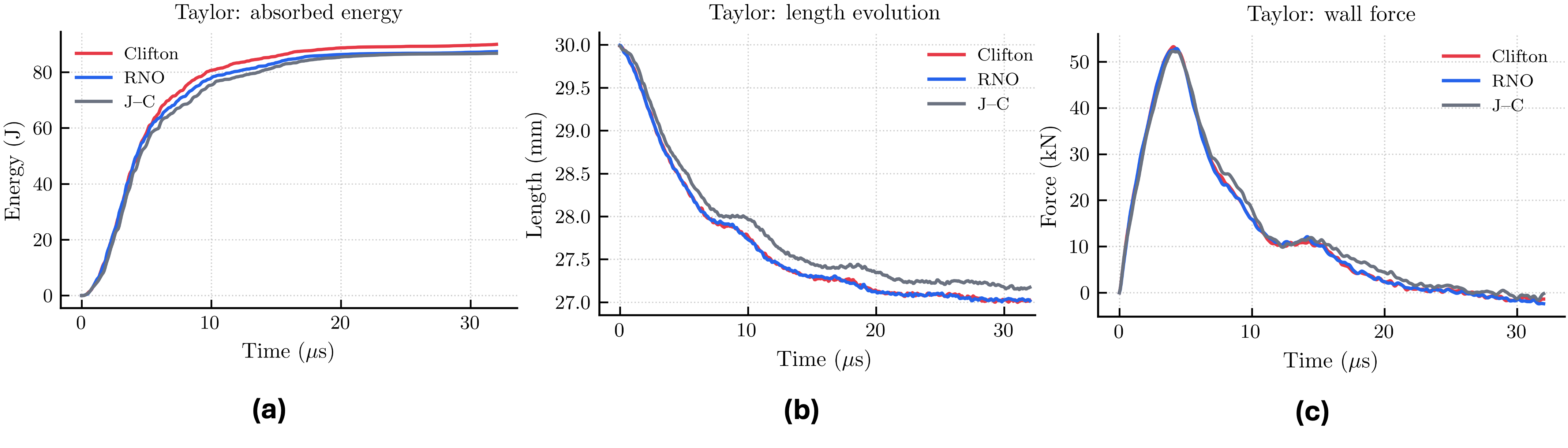}}
{\caption{The evolution of (a) absorbed energy, (b) length of impactor, and (c) wall force with time during the Taylor impact test.}\label{Fig:TaylorPlots}}
\end{figure}

In Figure \ref{Fig:TaylorPlots}, we plot the evolution of (a) absorbed energy, (b) axial length and (c) wall force with time. From the absorbed energy versus time plot (a), we see that all three models show a rapid increase in absorbed energy during the first few microseconds, corresponding to the wave-dominated loading phase when compressive stress waves are generated and propagate through the specimen. We see that the RNO atomistic surrogate constitutive model is in good agreement with both Clifton and Johnson-Cook models. From the plot of variation in axial length versus time (b), we see that following impact, all three models show a rapid decrease in length as compressive waves induce inelastic deformation. The RNO atomistic surrogate is in excellent agreement with Clifton's model, from the initial rapid shortening through the slower late-time evolution, indicating that it captures the correct effective stiffness, flow behavior, and rate dependence of polyurea. In contrast, the length calculated using the Johnson–Cook model is longer, which is consistent with a stiffer, rate-independent plastic response that lacks polymer-specific relaxation mechanisms. 

The wall reaction force shown in Figure \ref{Fig:TaylorPlots}c is a measure of the transfer of momentum at the impact interface. All three models show a sharp peak in force after the impact, corresponding to the arrival of the compressive wave at the anvil, followed by a decaying force as the specimen deforms and unloading waves develop. The RNO atomistic surrogate accurately reproduces both the peak force magnitude and the subsequent decay observed in Clifton's model, thereby correctly capturing both wave propagation and contact mechanics. In contrast, the force calculated using the Johnson–Cook model deviates slightly after the initial peak, during decay.


\subsection{Plate impact}
We perform a plate-impact test to evaluate each constitutive model for coupled stress-wave propagation, structural bending, and rate-dependent dissipation of a constrained polyurea plate under dynamic loading. Unlike the Taylor anvil test, which is predominantly uniaxial, the plate-impact configuration introduces a coupled response involving local indentation, flexural deformation, and repeated wave reflections from the clamped boundary \cite{meyers1994dynamic}.

\begin{figure}[h!]
\centering
{\includegraphics[keepaspectratio=true,width=1.0\textwidth]{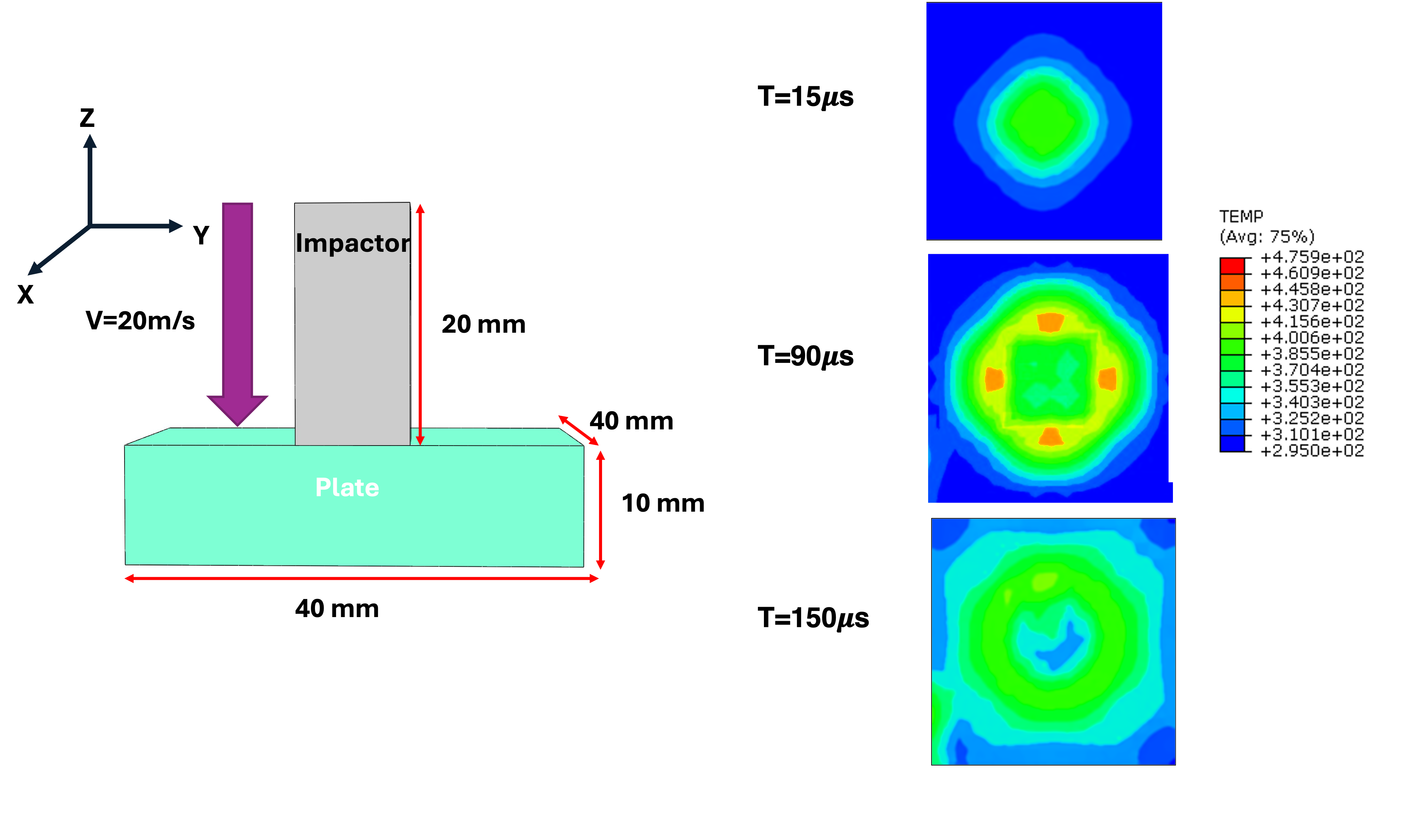}}
{\caption{The geometry of the plate impact test are schematically shown on the left. The evolution of temperature {(K)} across the midplane of the cylinder during the simulation is shown on the right.}\label{Fig:PlateTemp}}
\end{figure}

In our plate-impact test, a rigid $10 \times 10 \times 20$ mm projectile impacts the center of a $40 \times 40 \times 10$ mm polyurea plate. The lateral plate dimensions are selected so that the projectile footprint thereby reduces immediate boundary dominance and promotes a primarily local indentation response superimposed on global bending. The $10$ mm plate thickness provides sufficient stiffness to sustain stress-wave transmission through the thickness while avoiding excessive element distortion or contact-driven numerical instabilities in explicit dynamics. The projectile is assigned an initial velocity of $20$ m/s in the direction of impact. This velocity is chosen to generate measurable transient wave dynamics and rate-dependent deformation in polyurea, while remaining moderate enough to maintain robust explicit time integration and stable contact behavior for the selected mesh resolution and element formulation. One face of the plate is free and serves as the impact surface, while the opposite face is fully clamped. The contact between the projectile and the plate is modeled using hard normal contact with frictionless tangential behavior. The projectile is modeled as a solid with elastic modulus $E = 210$ GPa and $\nu=0.3$, ensuring that deformation of the projectile is negligible relative to the polyurea sample. The setup is shown in Figure \ref{Fig:PlateTemp}.

We use C3D8RT reduced-integration brick elements for the polyurea plate and C3D8R reduced-integration brick elements for the projectile, respectively\cite{abaqus2016theory,belytschko2013nonlinear}. The plate–projectile assembly contains $5,112$ elements and $6,284$ nodes, and the impact event is simulated in a single Abaqus/Explicit dynamic step of duration $300 \mu$s, sufficient to capture the transient impact phase and subsequent structural response.

In Figure \ref{Fig:PlateTemp}, we plot the contours of the temperature calculated using the RNO atomistic surrogate. At $t = 15\mu$s, the temperature rise is localized near the impact region, coinciding with the region of highest stress and strain-rate activity. In the constitutive implementation, temperature evolution is computed directly from local mechanical dissipation: the dissipated work at each integration point is converted into heat through a Taylor–Quinney-type relation, and the resulting temperature update contributes to the thermally softened material response. By $t = 90\mu$s, the heated zone grows radially outward, producing a broader annular distribution, consistent with expanding deformation and dissipation under peak loading. At $t=150 \mu$s, the temperature becomes more spatially distributed, indicating that the majority of heat generation has occurred during the peak-loading phase and that the system is transitioning into an unloading/relaxation regime with persistent residual heating in the region of accumulated deformation.

\begin{figure}[h!]
\centering
{\includegraphics[keepaspectratio=true,width=1.0\textwidth]{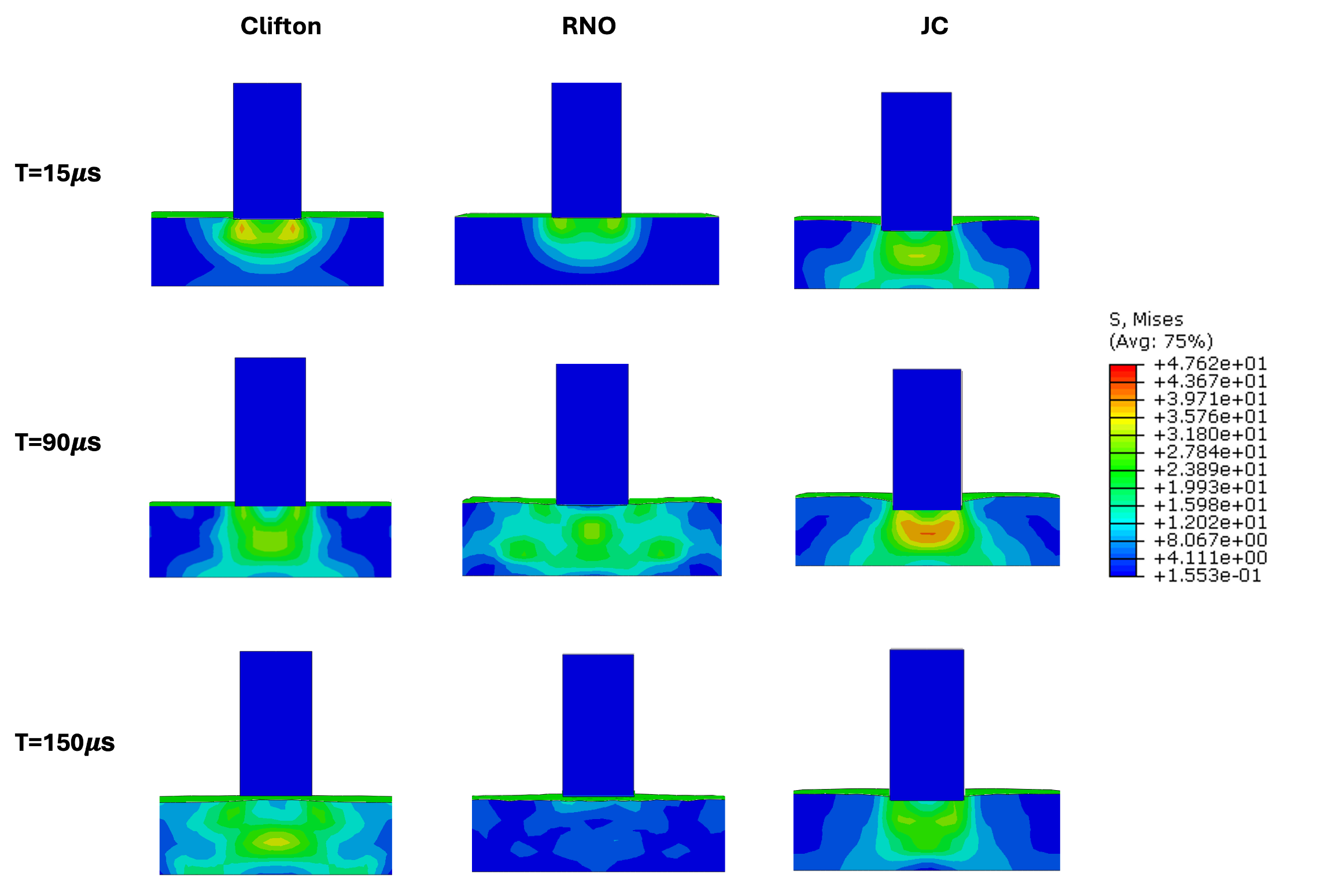}}
{\caption{Comparison of von Mises stress {(MPa)} for the three material models is shown at three different times for the plate impact test}\label{Fig:PlateStress}}
\end{figure}

Figure \ref{Fig:PlateStress} shows the contour plots of the von Mises stress during the plate impact test comparing the three constitutive models at $t=15$, $90$, and $150 \mu$s. At the early stage ($t=15 \mu$s), all three models show stress localization beneath the impacted area, corresponding to the onset of contact-induced compression and the initiation of stress waves propagating into the plate thickness. In Clifton's constitutive model and RNO atomistic surrogate, the high-stress region remains confined near the impacted surface with smooth spatial decay away from the impacted surface. The Johnson–Cook model shows a more concentrated stress core directly under the projectile, due to its stiffer, rate-independent plastic response. At $t=90 \mu$s, the von Mises stress is distributed over a larger area for all three models. Additionally, the maximum von Mises stress calculated with the RNO atomistic surrogate is in good agreement with Clifton's constitutive model. In contrast, the maximum von Mises stress calculated using the Johnson–Cook model is higher. At $t=150\mu$s, the maximum von Mises stress calculated using all three constitutive models is smaller compared to the previous time steps. This is due to the transition from loading-dominated deformation to unloading, stress-wave reflection from the clamped boundary, and post-impact relaxation. The maximum von Mises stress calculated using the RNO atomistic surrogate is smaller compared to Clifton's and Johnson-Cook constitutive models. Additionally, the Johnson–Cook model shows localized stress regions underneath the impact location, due to its inability to accurately capture stress relaxation and the absence of polymer-specific viscoelastic dissipation mechanisms.

\begin{figure}[h!]
\centering
{\includegraphics[keepaspectratio=true,width=1.0\textwidth]{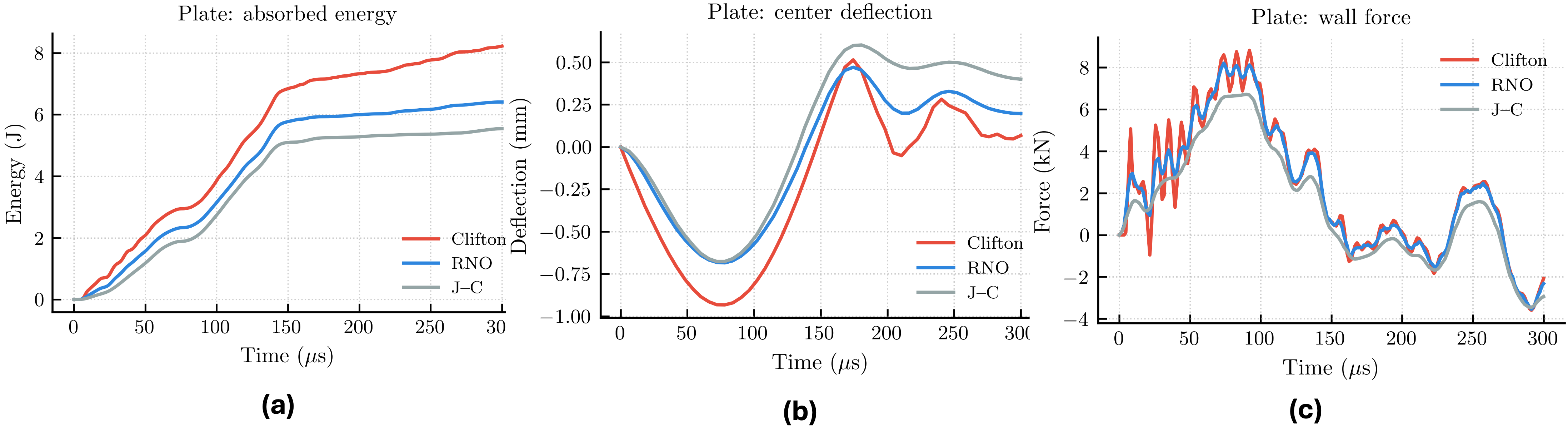}}
{\caption{The evolution of (a) absorbed energy, (b) deflection of the center of the plate, and (c) wall force with time during the plate impact test.}\label{Fig:PlatePlots}}
\end{figure}

Figure \ref{Fig:PlatePlots} shows the evolution of a) absorbed energy b) deflection of the center of the plate and c) wall force with time. The absorbed energy quantifies the conversion of work during impact into internal deformation and dissipation within the plate. For all three models, the absorbed energy increases monotonically throughout the event. Between $t = 0–150 \mu$s, the absorbed energy steeply increases with time, corresponding to active contact loading, stress-wave propagation, and rapid deformation beneath the projectile.  for $t> 150 \mu$s, the rate of increase in absorbed energy decreases for all three constitutive models, indicating reduced incremental dissipation as the system transitions away from peak loading. At all times, the absorbed energy predicted by Clifton’s model is the largest, followed by the RNO atomistic surrogate and, finally, the Johnson–Cook model. 

From the center deflection versus time plot, shown in Figure \ref{Fig:PlatePlots}b, we see that all three models show an initial downward deflection (negative), reaching a maximum magnitude during peak-loading ($t=70–100\mu$s), followed by unloading and rebound as contact forces diminish. The maximum deflection calculated using Clifton's constitutive model is the largest. At $t>150\mu$s, we see oscillations in center deflection due to reflection of the stress wave in the plate.

Figure \ref{Fig:PlatePlots}c shows the evolution of the wall reaction force with time. The wall reaction force quantifies the transfer of momentum and interaction between the contact and structure at the clamped boundary. From this figure, the wall force calculated by the RNO atomistic surrogate is in excellent agreement with Clifton’s constitutive model, including accurate capture of high-frequency oscillations, whereas the Johnson–Cook model yields a comparatively smoother, less oscillatory force response.

Overall, the plate-impact results demonstrate that the RNO atomistic surrogate captures the coupled stress-wave propagation, viscoelastic dissipation, and structural response of polyurea with high fidelity. Agreement with the experimentally derived Clifton viscoelastic model is observed in the evolution of temperature, stress distributions, absorbed energy, structural deflection, and wall reaction force. These comparisons indicate that the neural-operator surrogate preserves the essential history-dependent physics governing the dynamic deformation of polymeric materials. In contrast, the Johnson–Cook model systematically underpredicts dissipation, exhibits stiffer response during impact, and fails to reproduce the relaxation-driven redistribution of stresses characteristic of viscoelastic polymers. The results confirm that the RNO framework bridges the gap between atomistic fidelity and continuum-scale efficiency, enabling predictive multiscale simulations of rate-dependent polymer behavior in complex dynamic loading scenarios.

\section{Computational efficiency}\label{sec:efficiency}
In this section, we assess the computational efficiency of the atomistic-continuum multiscale framework presented in the paper. The motivation for introducing the recurrent neural operator(RNO) surrogate is to decrease the prohibitive computational cost of direct atomistic-continuum concurrent simulations. In a conventional concurrent multiscale (MD-FEM) framework, the macroscopic finite element solver requires a constitutive evaluation at every integration (quadrature) point and at every time increment. When the constitutive response is obtained from molecular dynamics (MD), a full microscopic simulation of a representative volume element (RVE) must be performed at each quadrature point and time step. We first {present details of computational resources used,} quantify the cost of an MD-FEM multiscale simulation without using a surrogate and then discuss the effort needed to develop the surrogate, followed by a discussion on the time taken by the multiscale simulations.

{All molecular dynamics simulations and neural-operator training were performed on the Oak Ridge Leadership Computing Facility Frontier supercomputer \cite{frontier2022} using GPU-accelerated compute nodes comprising one 64-core AMD EPYC processor and four AMD Instinct MI250X accelerators. The molecular dynamics simulations employed domain-decomposed parallel execution with GPU acceleration of force evaluation and neighbor-list construction, enabling efficient generation of the large ensemble of deformation trajectories required for training. The recurrent neural operator was implemented and trained in PyTorch using distributed GPU-based optimization and batched trajectory data resident on accelerator memory during backpropagation through time. All ABAQUS simulations were performed on a workstation equipped with an Intel Core i7 processor (8 physical cores, 16 logical cores), 32 GB RAM, and an NVIDIA GeForce RTX 2070 discrete GPU with 8 GB memory. The ABAQUS/Explicit simulations were executed primarily in CPU-parallel mode.}

{Next, we estimate the cost of MD-FEM simulations without the surrogate.} Let $N_{\mathrm{q}}$ denote the number of quadrature points and $N_{\mathrm{steps}}$ the number of timesteps in the macroscopic FEM solve. Let $T_{\mathrm{MD}}$ be the wall-clock time of a single MD solve at one quadrature point and time step, and let $T_{\mathrm{FEM}}$ denote the wall-clock time of one macroscopic FEM step excluding constitutive evaluations. The total wall-clock time $T_{{0}}$ of a fully concurrent MD-FEM simulation is
\begin{equation}
T_{\mathrm{0}} = N_{\mathrm{steps}}
\left(T_{\mathrm{FEM}} + N_{\mathrm{q}}\,T_{\mathrm{MD}} \right).
\label{eq:T2M}
\end{equation}

In the simulations considered here, the cyclic loading test contains {$672$} quadrature points, the Taylor impact problem contains $3928$ quadrature points, and the plate impact problem contains $5112$ quadrature points. Explicit dynamic analyses of stress-wave propagation and viscoelastic relaxation require on the order of $10^3$--$10^4$ time steps. Each MD simulation of the polyurea RVE requires $90$ seconds of wall-clock time on the OLCF Frontier exascale supercomputer \cite{frontier2022}. Consequently, the total number of MD calls scales as $\mathcal{O}(N_{\mathrm{q}} \times N_{\mathrm{steps}})$, leading to wall-clock times {of the order of $10^3$ hours on 1000 CPUs and GPUs}, rendering fully concurrent MD-FEM simulations computationally intractable for problems of this scale.

In our framework, each constitutive evaluation using MD simulation is replaced with a neural operator update. To distinguish the costs associated with data generation, training, and the multiscale simulation, we separate the workflow into an \emph{overhead} stage, involving data generation and training steps, and a \emph{simulation} stage with the RNO surrogate. Let $N_{\mathrm{train}}$ denote the
number of MD simulations used for training, $N_p$ be the number of parallel computing resources used for these simulations, and $T_{\mathrm{train}}$ be the wall-clock time taken to train the recurrent neural operator surrogate. Then, the wall-clock time taken to evaluate the overhead is
\begin{equation}
T_{\mathrm{overhead}} = \frac{N_{\mathrm{train}}}{N_p}\,T_{\mathrm{md}} + T_{\mathrm{train}} .
\label{eq:Toff}
\end{equation}
In this work, we have used $N_{\mathrm{train}}$ = $2000000$, $N_p$ = $1000$, and the times for MD simulations and training are, $T_{\mathrm{MD}}$ = $\approx90$ seconds, $T_{\mathrm{train}}$ = $\approx 1$ hour. Using these, we obtain the overhead time as $T_{\mathrm{overhead}}$ = $\approx 51$ hours. We note that though the total cost of generating the data is significant, it is performed only once for a given material and range of temperatures. To this end, the use of large-scale High Performance Computing resources for the purpose of data generation is critical, and increasing the number of computing resource will further decrease the wall-clock times. Furthermore, as each MD simulation is independent, they can be executed in embarrassingly parallel batches.  

\begin{table}[h!]
\centering
\caption{Wall-clock times (hours) of macroscale simulations}
\begin{tabular}{l c c c }
\hline
 Material model & Cyclic loading & Taylor impact & Plate impact \\
\hline
RNO atomistic surrogate & $2.9$ & $4.2$ & $4.7$ \\
Johnson-Cook & $2.2$ & $3.2$ & $3.7$ \\
Clifton &  $3.7$ & $5.2$ & $5.9$ \\
\hline
\end{tabular}
\label{tab:ExecutionTimes}
\end{table}

The wall-clock times for the three macroscale simulations ($T_{macro}$) using Finite Element Method with the three material models are shown in Table \ref{tab:ExecutionTimes}. From this, we see that the execution time with the Johnson-Cook material model is the smallest, and the time taken by the RNO atomistic surrogate is significantly smaller than the execution time using Clifton's material model.

\section{Concluding remarks and outlook}\label{sec:conc}
In this paper, we introduce a recurrent neural operator-accelerated concurrent multiscale framework that bridges the atomistic and continuum scales for history-dependent viscoelastic materials, with computational cost comparable to FEM simulations. To this end, we first develop a Recurrent Neural Operator (RNO) surrogate that learns the strain-history-to-stress operator from molecular dynamics simulations of viscoelastic materials and replaces prohibitively expensive fine-scale constitutive evaluations with the RNO surrogate in an explicit finite-element solver. The RNO surrogate is both temporally and spatially discretization independent, and encodes memory through learned latent internal variables. We have verified the accuracy and efficiency of the surrogate using multiscale simulations of polyurea and have benchmarked it against established constitutive models. Overall, the RNO surrogate is computationally efficient and accurately captures both the microscale and the multiscale response of polyurea, thereby enabling atomistic to continuum multiscale simulations of viscoelastic materials tractable.


Several extensions of the present framework merit further investigation. First, the construction of the neural-operator surrogate from atomistic data may be generalized to incorporate additional variables, including materials with temperature memory \cite{hollenweger2026temperature}, thermal conductivity, and rate-dependent dissipative mechanisms, thereby enabling a more complete treatment of coupled thermomechanical processes in polymeric systems. Second, the present study focuses on homogeneous atomistic training data obtained from molecular dynamics simulations of polyurea; future work will address heterogeneous microstructures and spatially varying material states by coupling the operator learning procedure with microstructure-resolved simulations or experimental datasets \cite{gupta2025designing}. From a numerical standpoint, adaptive strategies for enriching the training data during simulation, for example, through on-the-fly MD sampling or active learning, will improve predictive capability in regions of the deformation or temperature space not represented in the original dataset. Finally, the framework may be extended to other classes of history-dependent materials, including thermoviscoelastic polymers, elastomers, and soft composites, as well as to multiphysics settings where mechanical response couples to electromagnetic or chemical fields. Such developments would broaden the applicability of neural-operator-accelerated multiscale simulations for predictive analysis and design of complex material systems.

{We close by highlighting and interpreting the differences between the timescales accessible to molecular dynamics and those governing macroscopic material response. In the present study, the RNO is trained on MD trajectories spanning approximately $1 \mathrm{ns}$, which is sufficient to resolve the segmental relaxation modes of polyurea ranging from picoseconds to nanoseconds that dominate viscoelastic dissipation under the dynamic loading conditions considered here, ranging from tens to hundreds of microseconds. For materials whose macroscopic behavior is controlled by substantially slower processes, such as long-term physical ageing in amorphous silica, $\beta$-relaxations in polymer glasses, or slow shear transformations in metallic glasses occurring over timescales from minutes to years, the principal limitation is the training data. An RNO trained only on nanosecond-scale trajectories cannot recover relaxation modes that are absent from the sampled atomistic dynamics. A natural path forward is therefore to construct training datasets using temporally accelerated atomistic approaches capable of reaching substantially longer physical times than direct MD, including accelerated dynamics \cite{perez2009accelerated}, hyperdynamics \cite{voter1997hyperdynamics}, diffusive molecular dynamics \cite{li2011diffusive}. Recent efforts \cite{spinola2026predicting,saxena2023gnn} in using such techniques provide a promising direction for extending the present framework to materials governed by slow processes.}

\section*{Data availability}
The molecular dynamics simulation data used to train the model, together with the ABAQUS input files, are publicly available at the repository https://github.com/swarnavaghosh/RNOAtomisticToContinuum

\section*{Acknowledgements}
This research used resources of the Oak Ridge Leadership Computing Facility, a DOE Office of Science User Facility operated by the Oak Ridge National Laboratory under contract DE-AC05-00OR22725. 
This manuscript has been authored in part by UT-Battelle, LLC, under contract DE-AC05- 00OR22725 with the US Department of Energy (DOE). The publisher acknowledges the US government license to provide public access under the DOE Public Access Plan (http://energy.gov/downloads/doe-public-access-plan).

During the preparation of this work the authors used OpenAI (GPT-5) in order to search literature, correct grammar and revise language. After using this tool/service, the authors reviewed and edited the content as needed and takes full responsibility for the content of the published article.

\section*{Appendix A. True and predicted stress at 400 K and 500 K}\label{Sec:AppendixA}

\begin{figure}[h!]
\centering
{\includegraphics[keepaspectratio=true,width=1.0\textwidth]{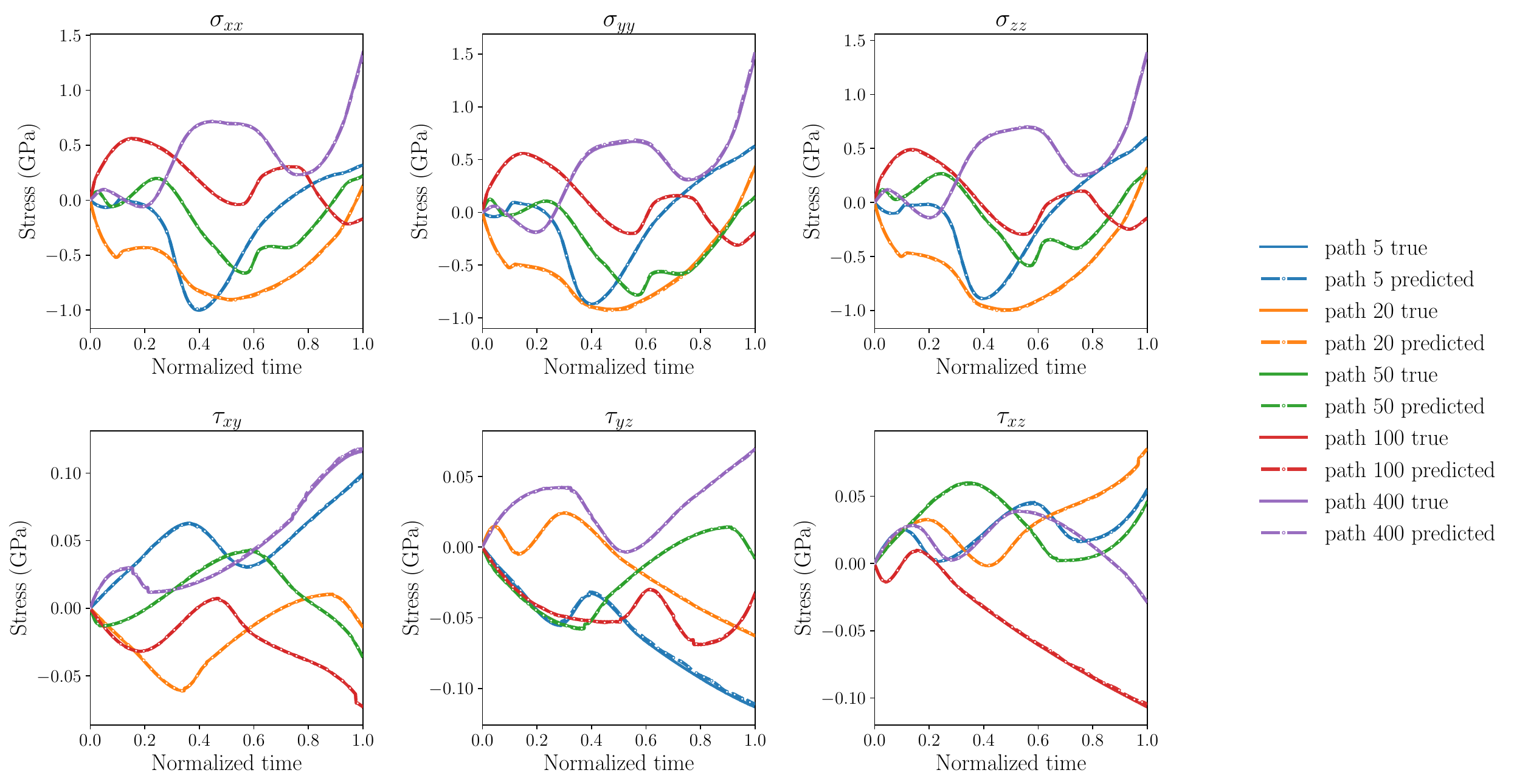}}
{\caption{Stress prediction and convergence for RNO at 400K.  The uniaxial stresses are shown in the top panel, and shear stresses are shown in the bottom panel.}\label{Fig:stress_400}}
\end{figure}

\begin{figure}[h!]
\centering
{\includegraphics[keepaspectratio=true,width=1.0\textwidth]{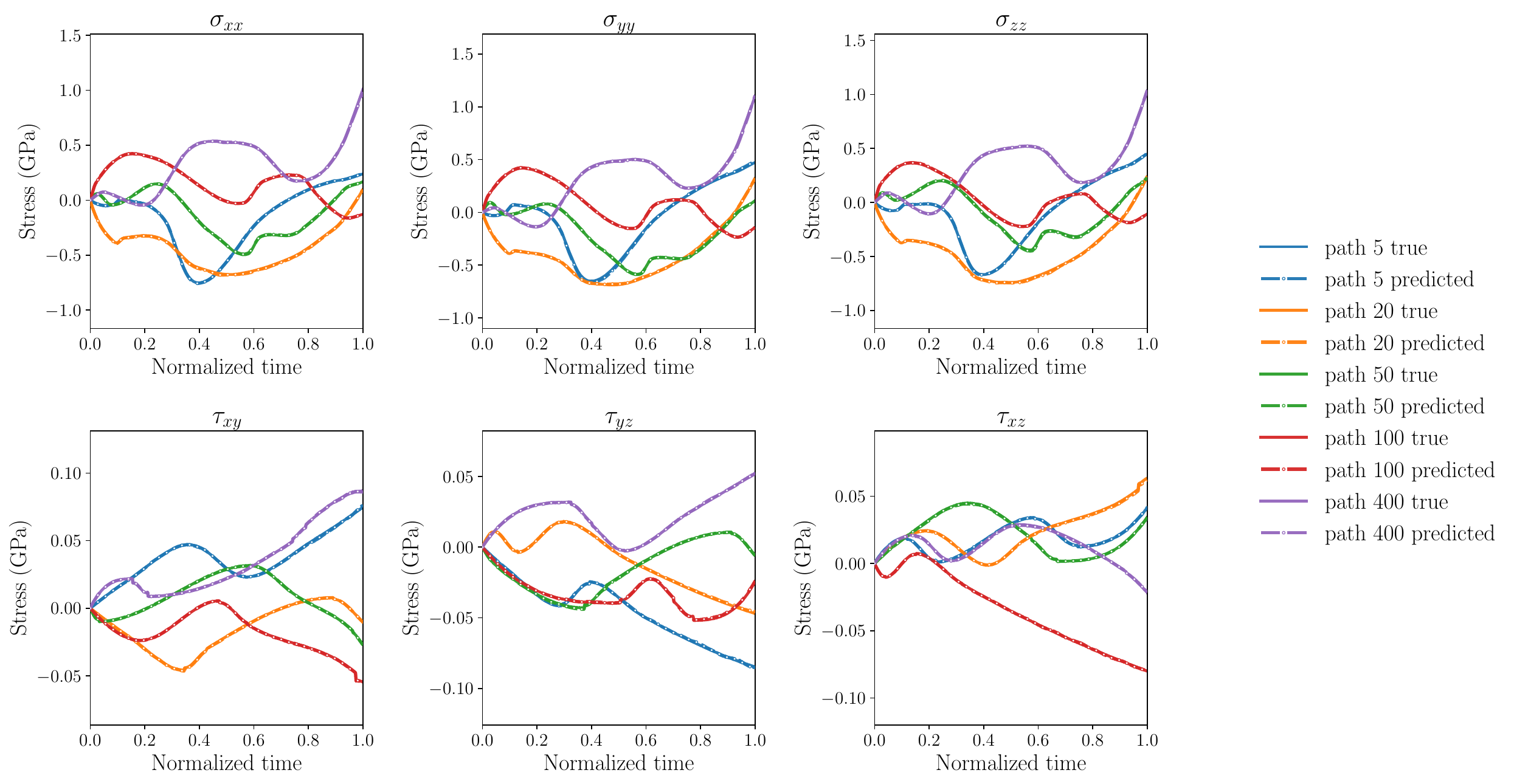}}
{\caption{Stress prediction and convergence for RNO at 500K.  The uniaxial stresses are shown in the top panel, and shear stresses are shown in the bottom panel.}\label{Fig:stress_500}}
\end{figure}

Figure \ref{Fig:stress_400} and \ref{Fig:stress_500}, shows the plots of true and predicted stress for temperatures $400$ K and $500$ K respectively.

\section*{Appendix B. Material models for assessment of atomistic surrogate}\label{Sec:AppendixB}
The atomistic surrogate in multiscale simulations is verified with a) an experimentally derived viscoelastic constitutive model of polyurea and b) Johnson-Cook material model. We give an overview of these models.

\subsection*{Experiment-derived viscoelastic constitutive model of polyurea}
Clifton and co-workers \cite{clifton2016physically} derived a physics based quasi-linear viscoelastic constitutive model for polyurea that is subjected to high pressures and strain rates, and calibrated against pressure--shear plate impact experiments for pressures of order $18$ GPa and strain rates ranging from $\dot{\varepsilon} \sim 10^{5}$--$10^{6}\,\mathrm{s^{-1}}$. Additionally, the instantaneous elastic response is described by finite-deformation isotropic hyperelasticity, in which the strain-energy density is multiplicatively decomposed as
\begin{equation}
W(J,\bar I_1,\bar I_2) = f(J)\,\hat W(\bar I_1,\bar I_2),
\end{equation}
where $J=\det \bm F$ is the Jacobian of the deformation gradient $\bm F$, and $f(J)$ is a Lennard--Jones--type volumetric function, $\bar I_1$ and $\bar I_2$ are the normalized invariants of the left Cauchy--Green tensor. In this decomposition, the shear modulus and shear-wave speed explicitly depend on the pressure, consistent with experimental observations. The distortional energy $\hat W$ is of the Mooney--Rivlin form,
\begin{equation}
\hat W = C_{10}(\bar I_1-3) + C_{01}(\bar I_2-3)+C_{00},
\end{equation}
where $C_{10}$, $C_{01}$ and $C_{00}$ are constants.
The numerical values of these constants are calibrated using experiments conducted by Clifton, Wang and Jiao \cite{clifton2016physically} and are given by $C_{10}=0.2684$ GPa, $C_{01}= 0.1098$ GPa, $C_{00}=0.2684$ GPa. Time-dependent effects are incorporated using a quasi-linear viscoelastic formulation in which the deviatoric stress $\bm{\sigma}$ relaxes from its instantaneous elastic value $\bm{\sigma}^e$ according to
\begin{equation}
\bm{\sigma}(t) = \bm{\sigma}^e(t) + \int_0^t \dot R(t-t')\,\bm{\sigma}^e(t')\,\mathrm{d}t',
\end{equation}
where $R(t)$ is a reduced relaxation function, constructed from a continuous distribution of relaxation times $t' \in [t_1,t_2]$,
\begin{equation}
R(t) = \varsigma + (1-\varsigma)\int_{t_1}^{t_2} e^{-t/t'}\,S(t')\,\mathrm{d}t',
\label{eq:R_of_t}
\end{equation}
where $\varsigma$ is the ratio of long-time to instantaneous shear modulus, the weighting function $S(t')$ is derived from a Boltzmann distribution of activation energies combined with an Arrhenius relation between relaxation time and energy barrier, leading to a logarithmically uniform distribution of relaxation times.

This viscoelastic constitutive model of polyurea is able to reproduce key features of measured particle velocity histories in both pressure–shear and symmetric pressure–shear plate impact experiments, particularly the delayed transverse response associated with shear stress relaxation~\cite{amirkhizi2006experimentally, clifton2016physically}. Relaxation of deviatoric stresses is essential for capturing the strong pressure-dependent shearing resistance of polyurea under high-rate loading, whereas the volumetric response is typically modeled as comparatively stiff with limited volumetric relaxation, consistent with the near-incompressible behavior of elastomeric polymers at high pressures.
Although the model achieves good agreement with experiments using a modest number of physically interpretable parameters, the dominant relaxation timescales are often comparable to the duration of the dynamic loading events. This motivates further validation of the constitutive description across broader temporal regimes to ensure predictive capability beyond the experimental timescales used for calibration.

\subsection*{Johnson and Cook}
The Johnson–Cook (J–C) constitutive model is one such widely adopted formulation for materials undergoing large strains, high strain rates, and temperature variations. It expresses the flow stress as a multiplicative combination of strain hardening, strain-rate sensitivity, and thermal softening, making it computationally efficient and well suited for explicit finite-element solvers such as Abaqus/Explicit. Due to its robustness and low computational cost, the J–C model is frequently used in impact and shock analyses where rate-dependent inelastic deformation dominates the response.
Although originally developed for metals, Taylor impact testing has been successfully extended to polymers and other soft materials to characterize high–strain-rate constitutive behavior \cite{sarva2007polycarbonate}. These experimental techniques provide the high-rate data necessary to calibrate phenomenological constitutive models used in impact simulations.

The JC equation is given below 
\begin{eqnarray}
\sigma = \left( A + B \, \varepsilon^{n} \right)
          \left( 1 + C \, \ln \frac{\dot{\varepsilon}}{\dot{\varepsilon}_0} \right)
          \left( 1 - T^{m} \right)
\end{eqnarray}

\begin{eqnarray}
    T = \frac{T - T_{\text{room}}}{T_{\text{melt}} - T_{\text{room}}}
\end{eqnarray}

\begin{table}[h!]
\centering
\caption{Johnson--Cook constitutive parameters \cite{choi2015polyurea}.}
\begin{tabular}{l c c}
\hline
Parameter & Symbol & Value \\
\hline
Yield stress coefficient & $A$ & 0.43 MPa \\
Hardening modulus & $B$ & 0.14 MPa \\
Strain hardening exponent & $n$ & 0.613 \\
Strain-rate sensitivity coefficient & $C$ & 1.61 \\
Thermal softening exponent & $m$ & 1.5 \\
Melting temperature & $T_m$ & 750 K \\
\hline
\end{tabular}
\label{tab:JC_polyurea}
\end{table}
In this work, the Johnson–Cook parameters summarized in Table~\ref{tab:JC_polyurea} were calibrated through curve fitting to Split Hopkinson Pressure Bar (SHPB) experimental data \cite{amirkhizi2006experimentally} for polyurea at strain rates up to $1.2\times10^4~\text{s}^{-1}$ \cite{choi2015polyurea}. These calibrated parameters were then implemented in Abaqus/Explicit to model the rate-dependent inelastic response of the polymer under high-strain-rate loading.

\section*{Author contributions}
Tanvir Sohail:  Conceptualization,  Methodology,  Software, Formal analysis,  Investigation, Writing - Original Draft,  Writing - Review \& Editing,  Visualization; 
Burigede Liu:  Conceptualization,  Methodology,  Software, Formal analysis,  Investigation, Writing - Review \& Editing;
Swarnava Ghosh:  Conceptualization,  Methodology,  Software, Formal analysis,  Investigation,  Writing - Original Draft,  Writing - Review \& Editing,  Visualization,  Supervision,  Project administration.

\section*{Conflict of interest declaration}
The author declares no competing interests.

\bibliographystyle{unsrt}
\bibliography{reference}
\end{document}